# Artificial Intelligence Across the Cardiac Amyloidosis Diagnostic and Management Pathway: From Single-Modality Detection Toward Multimodal Clinical Integration

**Diana Shadibaeva[1†], Rochak Dhakal[1†], Kui Zhang[2], Diana Paez[3], Xiaofeng Yang[4], Saurabh Malhotra[5,6], Weihua Zhou[1,7]***

[1] Department of Data Science, Michigan Technological University, Houghton, MI, USA

[2] Department of Mathematical Sciences, Michigan Technological University, Houghton, MI, USA

[3] Nuclear Medicine and Diagnostic Imaging Section, Division of Human Health, Department of Nuclear Sciences and Applications, International Atomic Energy Agency, Vienna, Austria

[4] Department of Radiation and Cellular Oncology, University of Chicago, Chicago, IL

[5] Division of Cardiology, Cook County Health and Hospitals System, Chicago, IL, USA

[6] Division of Cardiology, Rush Medical College, Chicago, IL, USA

[7] Center for Biocomputing and Digital Health, Institute of Computing and Cybersystems, and Health Research Institute, Michigan Technological University, Houghton, MI, USA

**† Diana Shadibaeva and Rochak Dhakal contributed equally to this paper.**

*** Corresponding author:**

**Weihua Zhou, Ph.D.**

**Department of Data Science, Michigan Technological University,**

**E-Mail: whzhou@mtu.edu**

***Abstract*** - Cardiac amyloidosis (CA) is increasingly recognized but remains substantially underdiagnosed because its clinical and imaging phenotype overlaps with more common cardiomyopathies. Definitive subtype assignment and management further require integration of multimodal evidence to distinguish transthyretin from light-chain disease. Machine learning and deep learning have been applied across the diagnostic and management pathway. These applications span electrocardiography (ECG), echocardiography, and health record-based case finding, as well as cardiac magnetic resonance (CMR) and nuclear interpretation, including single-photon emission computed tomography/computed tomography (SPECT/CT) biomarker quantification, prognostic modeling, and treatment response assessment. This narrative review synthesizes these studies by clinical tasks, namely screening, detection, quantification, prognosis, and longitudinal assessment after treatment initiation, rather than by input modality. This task-based organization clarifies why apparently similar AI models require different cohorts, reference standards, evaluation metrics, and implementation thresholds. The evidence reveals a maturity gradient. Binary detection and AI-assisted interpretation of cardiac scintigraphy with bone-avid tracers and SPECT/CT currently represent one of the more mature AI applications in cardiac amyloidosis, supporting standardized image interpretation and quantitative biomarker extraction. However, patient-level diagnosis still requires integration with monoclonal protein testing, SPECT/CT localization, clinical context, and biopsy or tissue typing when indicated. Detection is supported by large externally validated cohorts, and quantification by interpretable, outcome-linked measurement of myocardial tracer burden. By contrast, subtype-aware classification, prognostic risk stratification, and longitudinal assessment after treatment initiation remain at an early stage. These tasks are limited by small cohorts, enriched retrospective designs, heterogeneous labels, incomplete external validation, and uncertain calibration in realistic prevalence settings. Across tasks, high discrimination alone is insufficient. Clinically useful AI must demonstrate reliable performance against relevant mimics, maintain performance across patient subgroups, avoid circular reference standards, and improve referral or management decisions within existing diagnostic workflows. We argue that the field should now move beyond additional single modality classifiers toward multimodal, subtype-aware, and longitudinally validated systems that support, rather than replace, established diagnostic algorithms.

## 1. Introduction

Cardiac amyloidosis (CA) is a progressive infiltrative cardiomyopathy caused by extracellular deposition of misfolded amyloid fibrils within the myocardium, producing increased ventricular wall thickness, myocardial stiffness, diastolic dysfunction, restrictive physiology, arrhythmias, and eventual heart failure [1–3]. Most cases are caused by one of two major subtypes: transthyretin amyloidosis (ATTR) or immunoglobulin light-chain amyloidosis (AL). These subtypes differ fundamentally in pathogenesis, prognosis, and treatment urgency [2–4]. ATTR arises from misfolding of transthyretin protein and occurs as either wild-type ATTR, usually affecting older adults, or hereditary ATTR caused by pathogenic TTR variants. AL amyloidosis, by contrast, usually reflects an underlying plasma cell dyscrasia and follows a more rapidly progressive course requiring urgent hematologic evaluation [3,4] (Fig. 1). Therefore, identifying the subtype is not a classification exercise; it directly determines whether the patient requires hematology-directed therapy, ATTR-specific disease-modifying treatment, genetic testing, or tissue confirmation.

Although CA was historically viewed as rare, many studies and consensus statements emphasize that it is increasingly recognized but still remains underdiagnosed, especially among patients with heart failure with preserved ejection fraction (HFpEF), unexplained left ventricular hypertrophy (LVH), aortic stenosis, or other nonspecific cardiac abnormalities [2,3,5,6]. This increased recognition has been driven largely by improvements in noninvasive imaging, especially bone-avid tracer scintigraphy with 99mTc-PYP, 99mTc-DPD, and 99mTc-HMDP, as well as greater awareness of extracardiac red flags and disease-specific therapies [7–11]. Today, diagnosis relies on combining clinical suspicion with laboratory testing, electrocardiography (ECG), echocardiography, cardiac magnetic resonance imaging (CMR), nuclear scintigraphy, and, when needed, biopsy confirmation [2,7,9,12,13]. However, important challenges remain. Many patients are still diagnosed late, after substantial myocardial involvement has already occurred and subtype discrimination can remain uncertain in borderline or atypical cases [2,3,13]. The quantitative assessment of disease burden is not consistently reproducible across institutions; and widely accepted imaging markers of treatment response are still lacking [2,10,13]. These specific limitations have motivated sustained interest in machine learning (ML) and deep learning (DL) approaches to support CA diagnosis and management. ML and DL methods are well suited to problems that remain difficult in routine practice. These include identifying subtle disease-associated patterns in high-dimensional ECG, echocardiographic, and multimodal clinical data [14–16] reducing reader- and site-dependent variability during interpretation [17–19] and supporting automated differentiation of CA from phenotypically similar causes of increased left ventricular wall thickness such as hypertrophic cardiomyopathy (HCM) and hypertensive heart disease [17–20]. This complements emerging quantitative nuclear imaging approaches for assessing myocardial disease burden and potential longitudinal change, although validated imaging markers for treatment monitoring remain limited [21].

In recent years, ML and DL studies in CA have expanded across ECG [15,16,22], echocardiography [14,17–19], CMR [23–25], computed tomography (CT) [26], cardiac scintigraphy with bone-avid tracers [27,28], SPECT/CT [29–31], and structured clinical data [32,33], with many studies reporting strong discrimination in retrospective development or validation cohorts. Recent work has also begun to move beyond isolated single-modality diagnosis, including multimodal architectures combining echocardiography with electronic health record data (EHR) [32], parallel ECG-plus-echocardiography pipelines applied to preclinical disease trajectories [34], CMR-based deep learning for AL-versus-ATTR subtype discrimination [23], multicenter CMR machine learning for three-way classification among HCM , AL cardiac amyloidosis, and ATTR cardiac amyloidosis [25], and longitudinal AI-enabled quantification of treatment-response markers using SPECT/CT [31]. These studies show that the next stage of AI in CA is emerging, but they also highlight persistent limitations related to cohort size, external validation, and clinical integration. This review therefore complements recent work by comparing AI applications according to their intended clinical use and evidentiary maturity.

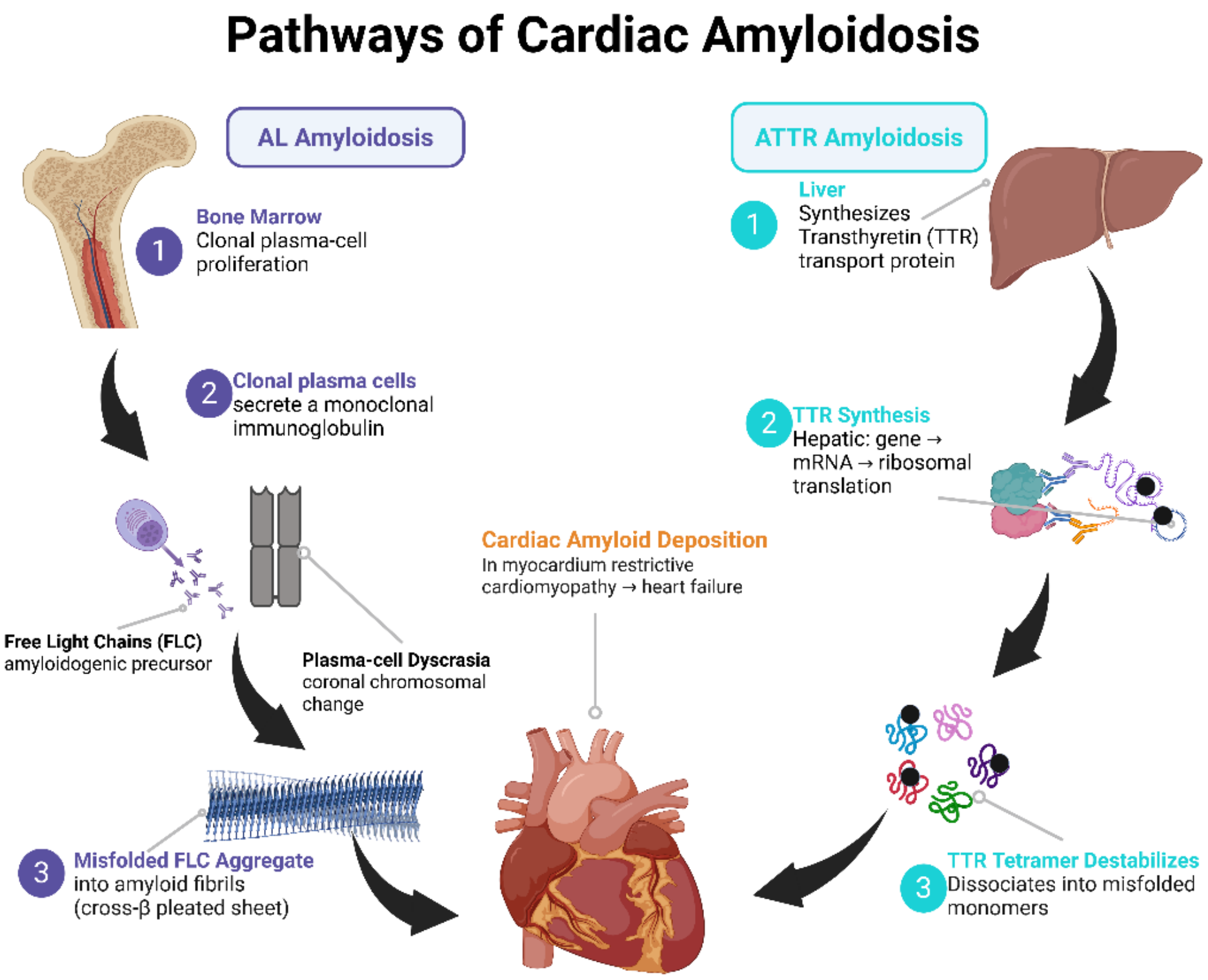

***Fig. 1: Pathways of AL and ATTR cardiac amyloidosis.*** *In AL amyloidosis, clonal bone-marrow plasma cells produce monoclonal immunoglobulin free light chains that misfold and aggregate into amyloid fibrils. In ATTR amyloidosis, liver-derived transthyretin becomes unstable, dissociates from tetramers into misfolded monomers, and aggregates into amyloid fibrils. In both pathways, extracellular myocardial amyloid deposition promotes restrictive cardiomyopathy and heart failure.*

Previous reviews have surveyed AI applications in CA largely by input modality, summarizing model architecture and performance metrics [35–38], while others have concentrated on specific modalities such as ECG and echocardiography [36]. The present review takes a different organizing principle. We synthesize the literature by clinical tasks including screening of at-risk populations, detection of suspected disease, quantification of disease burden, prognostic stratification, and treatment-response monitoring because each task imposes distinct requirements on dataset construction, label definition, evaluation strategy, and clinical integration. We critically appraise the evidence with attention to study design, sample size, external validation, class balance, and the independence of the reference standard. We compare evidence across modalities, with particular attention to clinical tasks in which AI has progressed beyond isolated diagnostic classification, including quantitative imaging, subtype-aware classification, and longitudinal assessment. The central argument of this review is that single-modality detection performance has matured as a research target, while multimodal, subtype-aware, and longitudinally validated models are beginning to emerge but remain limited by small cohorts, incomplete external validation, and weak integration with established diagnostic pathways. The next phase of clinically meaningful progress will depend on scaling, validating, and embedding these emerging approaches into real diagnostic workflows rather than producing additional single-modality detection studies.

## 2. Methods: Review Type and Search Strategy

This narrative review examines artificial intelligence applications across the cardiac amyloidosis diagnostic and management pathway, including screening, detection, subtype classification, disease-burden quantification, prognosis, and longitudinal assessment after treatment initiation. Targeted searches of PubMed/MEDLINE, Google Scholar, ClinicalTrials.gov, and reference lists were performed for studies published up to July 2026 using combinations of terms related to cardiac amyloidosis, ATTR-CM, AL amyloidosis, machine learning, deep learning, ECG, echocardiography, CMR, CT, PET, SPECT/CT, PYP, DPD, HMDP, prognosis, treatment response, multimodal AI, foundation models, federated learning, and explainable AI. We prioritized original studies that reported AI model development, validation, prospective or workflow implementation, or clinically interpretable quantitative biomarkers in cardiac amyloidosis. Preprints, trial registrations, and early implementation reports were considered only as emerging evidence and were not treated as equivalent to peer-reviewed multicenter studies. Articles were grouped by clinical tasks rather than input modality because screening, detection, quantification, prognosis, and longitudinal monitoring require different reference standards, evaluation metrics, thresholds, and implementation safeguards. Because this was a narrative review rather than a systematic review or meta-analysis, formal risk-of-bias scoring and pooled quantitative synthesis were not performed; however, we evaluated the literature qualitatively according to cohort realism, comparator selection, subtype confirmation, external validation, calibration, subgroup analysis, outcome linkage, and workflow readiness.

## 3. Clinical Background: Why the CA Diagnostic Pathway Needs AI?

### 3.1 The multimodal diagnostic workflow

Modern CA diagnosis is inherently multimodal. Patients with suspected CA are typically evaluated through clinical assessment, laboratory testing including monoclonal protein screening, ECG, and echocardiography. Depending on these findings, further evaluation may include CMR , bone-avid tracer scintigraphy with SPECT/CT, and biopsy when noninvasive testing is insufficient [2,7,10,39]. No single test is sufficient across all suspected cases or amyloid subtypes. Instead, each modality provides complementary information while also carrying specific limitations. A definitive diagnosis, particularly amyloid subtype assignment, therefore, depends on integrating evidence across the full diagnostic pathway. Fig. 2 summarizes this workflow and shows the diagnostic contribution of each modality. The following subsections describe each step in turn and highlight the specific gaps that have motivated the machine learning applications discussed in Section 4.

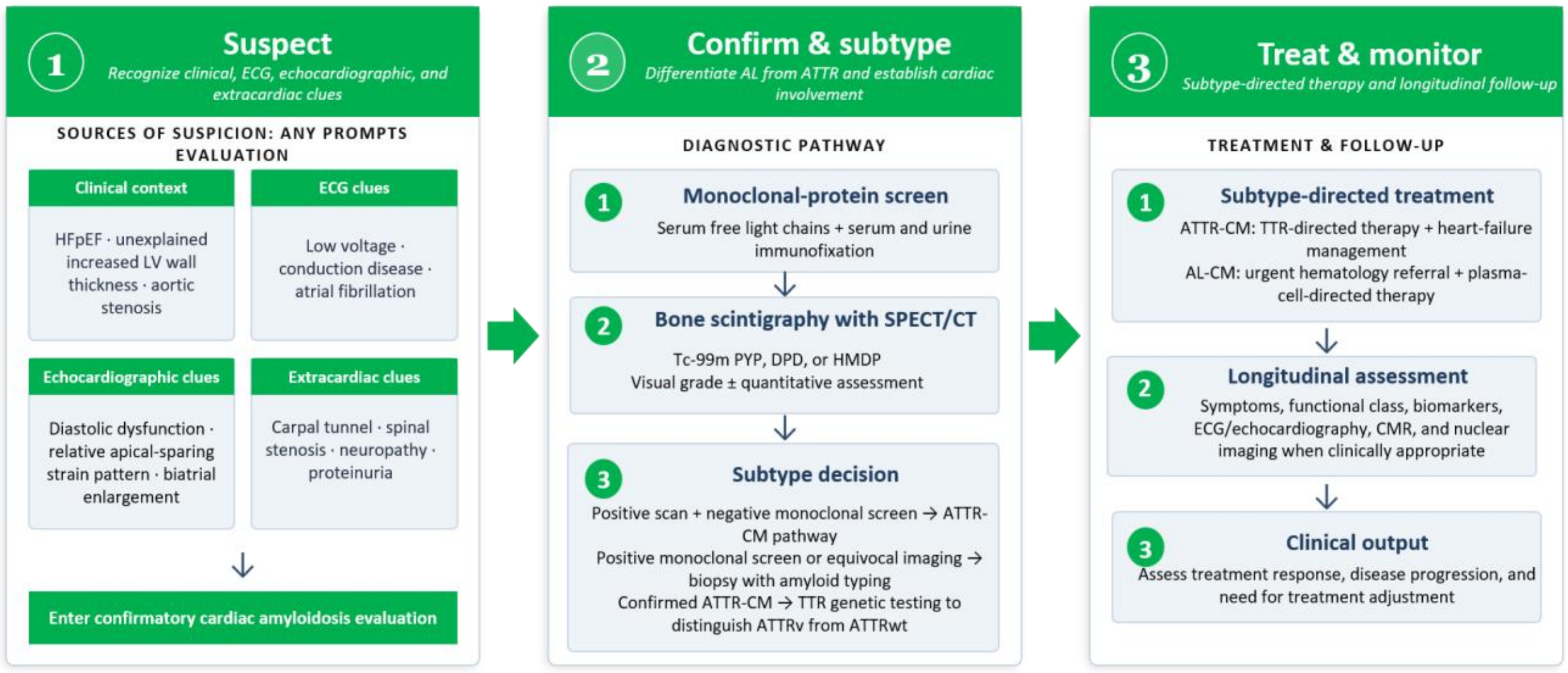

***Abbreviations:*** *AL-CM, light-chain cardiac amyloidosis; ATTR-CM, transthyretin cardiac amyloidosis; ATTRv/ATTRwt, variant/wild-type transthyretin amyloidosis; CMR, cardiovascular magnetic resonance; HFpEF, heart failure with preserved ejection fraction; sFLC, serum free light chains; TTR, transthyretin.*

***Fig. 2****: Established clinical workflow and diagnostic pathway for CA*

### 3.2 First-Line Evaluation: Clinical Assessment, ECG, and Echocardiography

Initial evaluation for suspected CA begins with clinical assessment, basic laboratory testing, ECG, and echocardiography. Suspicion may be raised by HFpEF, unexplained increased LV wall thickness, atrial arrhythmias, conduction disease, intolerance to standard heart-failure therapy, or extracardiac red flags such as carpal tunnel syndrome, lumbar spinal stenosis, spontaneous biceps tendon rupture, neuropathy, and autonomic dysfunction [3,13,39]. Once CA is suspected, monoclonal protein testing with serum free light chains, serum immunofixation, and urine immunofixation is essential because AL amyloidosis changes both diagnostic urgency and treatment pathway [2,13,39]. Cardiac biomarkers such as NT-proBNP and troponin support staging and prognosis but remain nonspecific [2–4,13]. Because these clinical clues are often subtle or attributed to more common cardiovascular disease, routine-data and multimodal AI models have been developed to identify at-risk patients earlier [32,33].

ECG is widely available and is often one of the first tests to suggest CA. Typical abnormalities include low QRS voltage, pseudoinfarct patterns, atrioventricular or intraventricular conduction delay, and atrial arrhythmias, reflecting myocardial infiltration and conduction-system involvement [2,13,39]. The classic mismatch between increased ventricular wall thickness on imaging and disproportionately low ECG voltage can be a useful clue, but it is neither sensitive nor specific. Many patients with ATTR cardiac amyloidosis do not show low-voltage patterns despite significant myocardial involvement [2,39]. This limitation has encouraged ECG-based machine learning and deep learning approaches that aim to detect subtle waveform patterns beyond routine visual interpretation [15,16,22,40].

Echocardiography is the main first-line imaging modality in suspected CA. Suggestive findings include increased left ventricular wall thickness, biatrial enlargement, diastolic dysfunction, restrictive filling physiology, valve thickening, right ventricular involvement, and reduced global longitudinal strain with relative apical sparing [2,9,20,39]. However, these findings overlap substantially with hypertensive heart disease, HCM, and aortic stenosis [2,20,39]. Therefore, echocardiography alone usually cannot confirm CA or determine subtype. This diagnostic uncertainty has motivated automated, multimodal, and video-based deep learning models designed to recognize CA-associated imaging patterns and improve referral for confirmatory testing [14,17–19,41].

### 3.3 Advanced Tissue Characterization with Cardiac Magnetic Resonance Imaging

CMR adds tissue characterization beyond what structural imaging can provide and has become an important tool in the evaluation of suspected CA [9,42,43]. Three CMR techniques are especially useful. Late gadolinium enhancement (LGE) can show diffuse subendocardial or transmural enhancement with abnormal myocardial nulling. Native T1 mapping can detect elevated relaxation times that reflect amyloid infiltration, while extracellular volume (ECV) quantification measures expansion of the interstitial space caused by amyloid deposition [9,42,43].

However, CMR cannot reliably distinguish ATTR from AL cardiac amyloidosis in isolation, because both subtypes can show LGE, elevated T1, and increased ECV, and subtype assignment still requires integration with monoclonal protein testing, scintigraphy, genetic testing, or biopsy when needed [2,13,43]. CMR is also more resource-intensive and less accessible in many settings. Its combination of rich tissue information and persistent subtype ambiguity makes it a natural target for AI models that can capture texture and spatial patterns beyond routine visual interpretation, including emerging approaches for AL-versus-ATTR discrimination [23,24,25,43,44].

### 3.4 Bone-Avid Tracer Scintigraphy and SPECT/CT for ATTR Cardiac Amyloidosis

Bone-avid tracer scintigraphy with 99mTc-PYP, 99mTc-DPD, or 99mTc-HMDP is central to the noninvasive diagnosis of ATTR cardiac amyloidosis [7,9–11,13]. In patients with negative monoclonal protein testing, grade 2 or 3 myocardial uptake, confirmed on SPECT or SPECT/CT, supports a non-biopsy diagnosis of ATTR cardiac amyloidosis [7,13] (Fig. 3).

Standard interpretation combines visual grading with SPECT or SPECT/CT localization [7,10,13,45]. Because planar imaging can be confounded by blood-pool activity, rib overlap, or adjacent focal uptake, SPECT or SPECT/CT is used to confirm myocardial localization [7,10,46]. Current nuclear interpretation remains partly dependent on visual grading, region-of-interest placement, and operator judgment, which can affect reproducibility [10,46,47]. In addition, planar semiquantitative indices such as the heart-to-contralateral-lung ratio can describe tracer uptake but do not capture the spatial extent or volumetric distribution of myocardial involvement [10,45]. Quantitative SPECT/CT metrics such as SUV, target-to-background ratio, volume of involvement, cardiac pyrophosphate activity, and cardiac pyrophosphate volume have therefore been proposed to better characterize myocardial tracer burden, although they are not yet standardized across institutions [21,30,46,48–50]. These challenges make nuclear imaging one of the most clinically promising areas for machine learning in CA. Automated uptake detection [27,28], anatomical localization and segmentation [29,30], and quantitative biomarker extraction [30,31,48,50] (Table 1) each address a specific part of the standardization problem.

PET amyloid imaging represents an emerging complementary nuclear approach in cardiac amyloidosis. Amyloid PET tracers may offer more direct quantification of myocardial amyloid burden and may have future value for diagnosis, subtype exploration, and longitudinal assessment, but current evidence remains limited by small cohorts, tracer heterogeneity, incomplete standardization, variable availability, and uncertain reimbursement. For this reason, PET is best framed as promising but not yet part of routine ATTR-CM diagnostic algorithms in most settings.

**Table 1. Quantitative biomarkers of myocardial amyloid burden on bone-avid tracer scintigraphy and SPECT/CT.**

| Biomarker | Full name | What it captures | Typical interpretation | References |
|---|---|---|---|---|
| Perugini grade | Visual grading scale, 0–3 | Categorical comparison of myocardial uptake to rib uptake | Grade 2 uptake, equal to rib uptake, or grade 3 uptake, greater than rib uptake, is considered visually positive; in the absence of monoclonal protein and with myocardial localization confirmed on SPECT or SPECT/CT, grade 2–3 uptake supports non-biopsy ATTR-CA diagnosis | [7,10,13,45] |
| H/CL ratio | Heart-to-contralateral-lung ratio | Semiquantitative ratio of cardiac to contralateral lung activity on planar imaging | Semiquantitative planar index of cardiac tracer uptake, used mainly for 99mTc-PYP; may help describe uptake burden and follow serial change over time. Tracer-, timing-, protocol-, and ROI-dependent. | [10,45] |
| SUVmax | Maximum standardized uptake value | Peak voxel uptake intensity within a defined myocardial region | Higher values reflect greater tracer accumulation, but SUVmax is sensitive to single-voxel noise and reconstruction differences | [46] |
| SUVpeak | Peak SUV within a small reference volume | Averaged peak uptake intensity over a small standardized volume | Less sensitive to noise than SUVmax; reported to correlate with Perugini visual score in quantitative DPD SPECT/CT studies | [46] |
| TBR | Target-to-background ratio | Ratio of myocardial uptake to a defined background reference region | Continuous measure of relative myocardial uptake; reported in Miller et al. 2024 to show strong separation between ATTR-CA and non-ATTR-CA groups but best framed as a quantitative uptake marker rather than a standalone diagnostic criterion. | [30] |
| VOI | Volume of involvement | Volume of myocardium with above-threshold tracer activity | Continuous measure of the spatial extent of myocardial involvement; reported in Miller et al. 2024 to show strong separation between ATTR-CA and non-ATTR-CA groups and may support disease-burden assessment. | [30] |
| CPA | Cardiac pyrophosphate activity | Composite measure combining uptake intensity and involved volume | Continuous measure reflecting intensity × extent of abnormal PYP uptake; reported in Miller et al. 2024 to show strong group-level separation and was associated with cardiovascular death or HF hospitalization after age adjustment. | [30] |
| CPV | Cardiac pyrophosphate volume | Volumetric measurement of abnormal myocardial PYP uptake on SPECT/CT | Continuous measure of tracer-burdened myocardial volume; reported to correlate inversely with LVEF and positively with posterior wall thickness and QRS duration | [49] |
| Quantitative SUV-based DPD measures | Standardized uptake value and normalized SUV measures | Absolute or normalized DPD uptake intensity on SPECT/CT | Quantitative DPD uptake is feasible and correlates with Perugini visual score, but thresholds and acquisition protocols remain insufficiently standardized | [46] |

*Abbreviations: ATTR-CA = transthyretin cardiac amyloidosis; CPA = cardiac pyrophosphate activity; CPV = cardiac pyrophosphate volume; DPD = 99mTc-3,3-diphosphono-1,2-propanodicarboxylic acid; H/CL = heart-to-contralateral-lung; HF = heart failure; LVEF = left ventricular ejection fraction; PYP = 99mTc-pyrophosphate; SPECT/CT = single-photon emission computed tomography/computed tomography; SUV = standardized uptake value; TBR = target-to-background ratio; VOI = volume of involvement.*

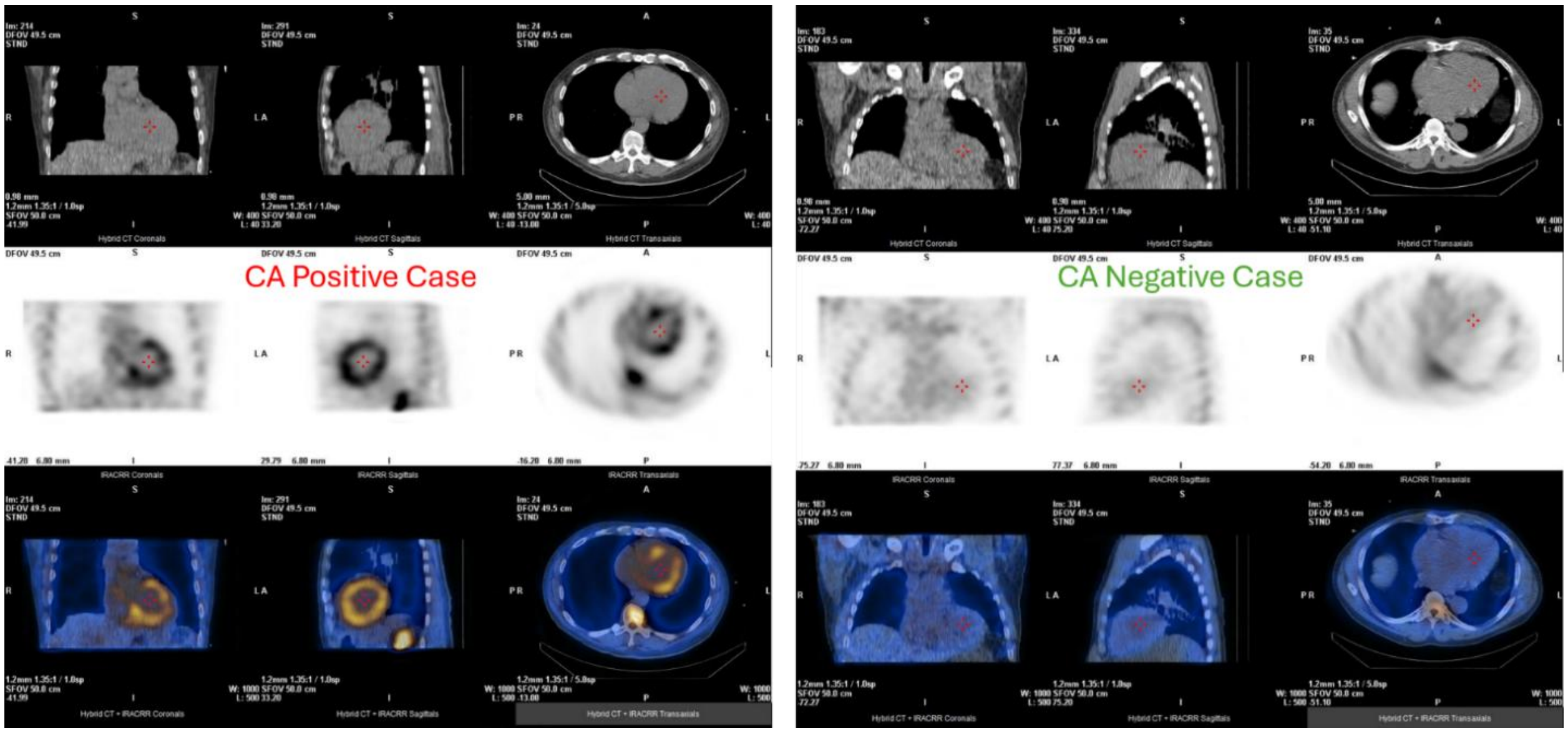

***Fig. 3**: **Representative positive and negative PYP SPECT/CT patterns for cardiac amyloidosis.** The positive case shows diffuse left ventricular myocardial tracer uptake comparable to or greater than rib activity, whereas the negative case shows no significant myocardial retention, with activity largely limited to physiologic rib background.*

### 3.5 Subtype Determination, Biopsy, and the Implications of Disease-Modifying Therapy

Distinguishing ATTR from AL cardiac amyloidosis is essential because the subtypes have overlapping imaging findings but differ in pathogenesis, prognosis, and treatment [2,4,13]. Subtype evaluation begins with monoclonal protein testing using serum free light chains, serum immunofixation, and urine immunofixation [2,13,39]. When monoclonal protein testing is abnormal, AL amyloidosis must be excluded before ATTR cardiac amyloidosis is diagnosed, even if bone-avid tracer uptake is present [2,7,13]. Histologic confirmation with amyloid typing remains the reference standard, and endomyocardial biopsy is considered when AL amyloidosis cannot be excluded, nuclear imaging is equivocal, or the clinical presentation is atypical [2,13,39].

The emergence of subtype-specific therapy has increased the importance of accurate and timely subtype determination. AL amyloidosis requires urgent hematology-directed treatment, whereas ATTR cardiac amyloidosis may be treated with ATTR-specific disease-modifying therapy [4,13]. This creates two AI-relevant gaps: imaging-based ATTR-versus-AL discrimination remains incomplete [13,23,25], and objective imaging markers of amyloid burden and longitudinal change have only recently begun to emerge [31,34].

Contemporary ATTR-CM management is also evolving rapidly with transthyretin stabilizers and silencers. Tafamidis remains the established reference therapy, while acoramidis and vutrisiran have expanded the disease-modifying treatment landscape in selected patients [51]. This therapeutic evolution increases the clinical importance of earlier diagnosis, subtype certainty, and validated longitudinal markers, but it does not yet establish SPECT/CT, PET, CMR, or echocardiographic AI markers as regulatory-grade treatment-response endpoints [52].

## 4. Artificial Intelligence Across the Cardiac Amyloidosis Diagnostic Pathway

Current cardiac amyloidosis diagnosis is limited by late detection, protocol-dependent interpretation, incomplete disease-burden quantification, and difficulty distinguishing ATTR from AL using imaging alone. These gaps motivate ML and DL approaches that can support earlier screening, automate interpretation, extract quantitative imaging biomarkers, and provide more objective markers for disease progression and treatment response.

### 4.1 Screening: Identifying Patients Who Require Further Evaluation

In CA screening, AI models are used to identify patients who may have unrecognized cardiac amyloidosis and should undergo further evaluation. These models typically use widely available data such as ECG, echocardiography, structured clinical variables, laboratory results, or EHR records. Screening is most relevant in patients with clinical features that raise suspicion for CA, including HFpEF, unexplained LVH, aortic stenosis, or nonspecific cardiac abnormalities in older adults.

#### 4.1.1 Electrocardiography Screening

The ECG is widely available, inexpensive, and obtained early in the clinical pathway, making it a natural target for AI-based CA screening. Schrutka et al. showed CA-associated electroanatomical signatures could be translated into a surface-ECG-based diagnostic algorithm using machine learning [16]. Grogan et al. developed an AI-ECG model using ECGs from patients with AL or ATTR cardiac amyloidosis and age- and sex-matched controls, reporting an AUC of 0.91. Its reduced-lead models also performed well, including single-lead V5 and 6-lead models with AUCs of 0.86 and 0.90, respectively. Importantly, among patients with ECGs available before diagnosis, the model identified CA more than six months before clinical diagnosis in 59% of cases [15]. Harmon et al. evaluated the same AI-ECG algorithm in 440 CA patients and 6,600 age- and sex-matched controls, reporting an AUC of 0.84. Performance was lower in specific subgroups, including patients with LVH, left bundle branch block, and Hispanic patients, with the Hispanic subgroup showing the lowest reported AUC of 0.66 [22]. Goto et al. extended the field by developing a fully automated, multi-institution pipeline using ECGs and echocardiograms as inputs. For the ECG model alone, performance was strong across institutional test sets, with ECG C-statistics of 0.91, 0.85, and 0.86; however, simulated deployment showed low PPV of approximately 0.03–0.04 at 0.52–0.71 recall [14]. Haimovich et al. evaluated AI-enabled ECG classification across 50,709 patients with LVH-associated diseases, including 304 with cardiac amyloidosis, and reported an AUROC of 0.95 for CA. Importantly, the cohort included clinically relevant hypertrophic phenotypes such as HCM , hypertension, and aortic stenosis, supporting evaluation beyond simple CA-versus-control classification [53]. Vrudhula et al. showed that CA screening performance depends strongly on case definition, control selection, and test population. Using approximately 1.3 million ECGs from 341,989 patients, they found that model AUCs ranged from 0.660 to 0.898 in matched held-out test sets, but from 0.467 to 0.898 when evaluated in a broader Cedars-Sinai general patient population cohort [40]. González-López et al. evaluated the Willem AI platform for ATTR-CM detection using 20,754 ECGs from 2,901 individuals, reporting a test AUC of 0.88, sensitivity of 80.7%, and specificity of 78.5% [54]. Collectively, these studies show that AI-ECG can raise suspicion for CA or ATTR-CM, but performance remains highly dependent on disease definition, population composition, and deployment prevalence.

### 4.1.2 Echocardiography Screening

Echocardiography is widely used in patients with heart failure, increased wall thickness, aortic stenosis, and other settings where CA may be missed, making it a common platform for AI screening. Approaches include measurement-based models, multiparametric scores, and video-based deep learning models trained on routine echocardiographic clips.

Chang et al. applied a random forest model to 19 routine echocardiographic parameters from 3,603 studies in 636 patients, identifying CA with an AUC of 0.84, sensitivity of 0.82, and specificity of 0.73; important predictors included mitral A-wave velocity, global longitudinal strain, LV posterior wall thickness, and left atrial area [55]. Goto et al. developed fully automated ECG- and echocardiography-based CA detection models. The apical four-chamber echocardiography video model achieved a C-statistic of 0.96 internally and 0.91, 0.89, 1.00, and 0.96 across external cohorts. The model also distinguished CA from clinically relevant hypertrophic mimics, including HCM, hypertensive heart disease, and end-stage renal disease, with C-statistics ranging from 0.87 to 0.96 [14]. Duffy et al. introduced EchoNet-LVH for automated LVH phenotyping, including LV geometry measurement and classification of LVH etiologies such as CA and HCM [17]. In international validation, EchoNet-LVH detected CA from parasternal long-axis and apical four-chamber videos with an AUC of 0.896 in 574 CA patients and 979 controls [18]. Ioannou et al. compared an AI-derived multiparametric echocardiographic score with a fully automated video model across 5,776 patients, showing that both approaches could identify CA, while the video model achieved stronger diagnostic performance and classified a larger proportion of patients [19]. Slivnick et al. proposed a workflow-compatible single-clip strategy for CA detection from one apical four-chamber video. In external validation across 18 sites, the model distinguished 597 CA patients from 2,122 controls, including HFpEF, hypertensive LVH, obstructive and nonobstructive HCM, and suspected CA cases excluded by Tc-PYP scintigraphy; after excluding uncertain AI outputs, it achieved an AUROC of 0.93, sensitivity of 0.85, and specificity of 0.93 [41]. Together, these studies show that echocardiography-based AI has progressed toward workflow-compatible screening using automated measurement extraction, end-to-end video analysis, and single-clip models.

### 4.1.3 Routine-Data, Longitudinal, Multimodal, and Workflow-Based Screening

Not all AI-based CA screening requires dedicated imaging interpretation. Several studies have explored routine laboratory values, claims- and EHR-derived variables, longitudinal ECG and echocardiographic signals, and staged multimodal workflows. Agibetov et al. provided early proof-of-concept that routine laboratory parameters could distinguish CA-related heart failure from non-CA heart failure [56]. Pan et al. extended this strategy using an XGBoost model based on routine blood-test variables in 6,563 patients with LVH, including 261 patients with CA. In internal validation, the model achieved an AUC of 0.95, sensitivity of 0.92, and specificity of 0.95 [33]. Huda et al. used medical claims and ICD-code patterns to identify patients at risk for wild-type ATTR-CM. Model performance decreased from an AUROC of 0.93 in the internal test setting to 0.80 when applied to a single-center real-world EHR heart-failure cohort [57]. This model has since been evaluated in real-world settings, including implementation at an academic medical center and application to a national hospitalized heart-failure registry [58,59].

Recent studies have also moved beyond single-time-point prediction. Oikonomou et al. assessed AI-enabled ECG and echocardiography in patients referred for nuclear cardiac amyloid testing. Across 7,352 echocardiograms and 32,205 ECGs, AI-derived ATTR-CM probabilities diverged as early as three years before nuclear testing. A double-negative AI-ECG/AI-echo screen provided high sensitivity, whereas a double-positive screen provided higher specificity across cohorts [34]. Feng et al. combined echocardiographic videos with structured EHR variables in a transformer-based multimodal model. The model achieved an AUROC of 0.94 using 5-fold cross-validation, although the study was preliminary because it included only 41 patients [32].

Workflow-based approaches have also been evaluated. Goto et al. evaluated a staged screening workflow in which ECG-AI was used as a prescreening step before echocardiography-AI. In simulated deployment, this sequential strategy increased the estimated PPV of echo-based screening from approximately 0.33 to 0.74–0.77 [14]. ATTRACTnet combined ECG waveforms, echocardiographic measurements, demographics, and diagnosis codes for amyloidosis-associated orthopedic conditions, achieving AUROCs of 0.85 internally and 0.82 externally. In a prospective clinical workflow, 24 of 50 tested AI-flagged patients were diagnosed with ATTR-CM [60]. TRACE-AI is an ongoing multicenter observational study using routine ECG, point-of-care ultrasound, and transthoracic echocardiography to estimate ATTR-CM underdiagnosis across diverse health systems. It should be considered an emerging deployment study until diagnostic-yield and outcome results are available [61].

#### 4.1.4 Overall Interpretation of Screening Evidence

The screening literature shows that AI models for CA have been developed across routine data sources, including ECG, echocardiography, laboratory values, claims data, and EHR-derived variables. Their primary role is not definitive diagnosis, but risk enrichment: identifying patients who may benefit from confirmatory evaluation. The most clinically informative studies are those tested in realistic comparator populations, including HFpEF, hypertensive LVH, HCM, aortic stenosis, and suspected CA cases later excluded by confirmatory testing. Across studies, AUC alone is insufficient; screening performance must also be interpreted in relation to disease prevalence, comparator selection, PPV, calibration, and the burden of unnecessary downstream testing.

### 4.2 Detection: Diagnostic Support in Patients with Established Suspicion

Detection models operate downstream of screening, clinical suspicion, or referral for confirmatory evaluation. Unlike screening models, which aim to identify patients who may need further testing, detection models support interpretation in patients who already have a suggestive phenotype or dedicated amyloid imaging study. In this setting, AI is used to automate recognition of CA-suggestive findings, reduce interpretive variability, and support diagnostic adjudication.

#### 4.2.1 Cardiac Scintigraphy with Bone-Avid Tracers Detection

Bone-avid tracer scintigraphy is one of the most developed single-modality detection areas for CA AI. However, the clinical target should be stated carefully. Most models detect abnormal cardiac uptake suggestive of CA or ATTR-CM; they do not replace the full diagnostic algorithm, which also requires exclusion of monoclonal protein disease and appropriate confirmatory interpretation.

Earlier, Halme et al. trained CNNs on 1,334 HMDP bone-scintigraphy studies to detect and classify cardiac uptake according to Perugini grade. The models achieved AUCs of at least 0.85 for four-class Perugini grading and greater than 0.88 for distinguishing negative from positive uptake, demonstrating early feasibility of automated uptake classification [62]. Delbarre et al. developed a convolutional neural network to detect significant cardiac uptake on whole-body technetium-99m cardiac scintigraphy with bone-avid tracers using image-level labels, with Perugini grade ≥2 as the positive class [27]. The model achieved high performance in both internal and external testing, with AUCs of 0.999 and external sensitivity and specificity of 0.961 and 0.995, respectively [27]. Spielvogel et al. extended this evidence in a large international multicenter study of 16,241 patients and 19,401 scans from nine centers in four countries, using multiple tracers including DPD, HMDP, MDP, and PYP [28]. Their model achieved AUCs of 1.000 in Austrian cross-validation and 0.997, 0.925, and 1.000 in independent United Kingdom, Chinese, and Italian cohorts, respectively [28]. Salimi et al. further supported this direction with a fully automated deep learning pipeline for ATTR-CM detection and scoring on whole-body scintigraphy using multi-tracer, multi-scanner, and multicenter data [63]. Mo et al. developed a SPECT/CT radiomics model in 290 patients with suspected CA who underwent PYP or DPD imaging. A stacking ensemble achieved AUCs of 0.871 in validation and 0.824 and 0.839 in two external test cohorts, although this preprint should be interpreted as emerging rather than definitive evidence [64].

Together, AI studies of cardiac scintigraphy with bone-avid tracers show strong performance for detecting CA- or ATTR-CM-suggestive myocardial uptake, especially in multicenter and multi-tracer datasets. However, because most models are evaluated against expert imaging interpretation or imaging-derived labels, their current role is best framed as standardized AI-assisted uptake detection, rather than replacement of patient-level diagnostic adjudication.

#### 4.2.2 Cardiac Magnetic Resonance Imaging-Based Detection

CMR provides detailed tissue characterization and can capture CA-related abnormalities in ventricular morphology, function, late gadolinium enhancement distribution, and myocardial texture. Compared with cardiac scintigraphy with bone-avid tracers, CMR-based detection is a more heterogeneous ML problem because CA findings can overlap with HCM, hypertensive LVH, aortic stenosis, and other infiltrative or restrictive cardiomyopathies.

Martini et al. developed one of the early deep learning approaches for CA detection using CMR in 206 patients with suspected disease [44]. Their model, applied to LGE images, achieved an AUC of 0.982 for detecting CMR patterns consistent with CA. A conventional ML model using structured CMR features also performed strongly, with an AUC of 0.952, and was not statistically different from the deep learning approach [44]. Agibetov et al. later evaluated CNNs for automated CA diagnosis from CMR in 502 patients, including 82 with CA. The best fine-tuned CNN achieved an AUC of 0.96, sensitivity of 0.94, and specificity of 0.90 in cross-validation [65].

Radiomics studies support a complementary direction. Zhou et al. showed that LGE-CMR radiomic features provided incremental diagnostic information beyond visual assessment and conventional quantitative CMR parameters [66]. Jiang et al. focused on differentiating CA from HCM using non-contrast cine-CMR radiomics, which is clinically relevant because HCM is an important phenotypic mimic [45]. More recently, Cockrum et al. applied a vision transformer architecture to cine and LGE CMR images for CA detection, expanding the evidence base beyond conventional CNNs [24]. Eckstein et al. evaluated a complementary non-contrast approach using CMR-derived bi-atrial and right-ventricular strain and functional features in 43 patients with CA, 20 with HCM, and 44 healthy controls. A radial-basis-function support vector machine achieved 0.90 accuracy and an AUC of 0.996, suggesting that engineered multi-chamber functional features may also support CA detection without relying on contrast-enhanced tissue characterization [67].

CMR-based AI has also been investigated for ATTR-versus-AL subtype classification. Germain et al. trained VGG16 CNNs on cine-CMR and LGE images from 120 patients, including 70 with AL and 50 with ATTR cardiac amyloidosis [23]. Patient-level AL-versus-ATTR accuracy was 0.750 for cine-CNN and 0.611 for LGE-CNN, compared with 0.617–0.675 for experienced readers; corresponding AUCs were 0.839, 0.679, and 0.714 for the best human reader [23]. Weberling et al. evaluated subtype-aware classification in a larger retrospective multicenter, multivendor CMR cohort including 400 participants: 95 healthy volunteers, 94 patients with HCM, 95 with AL cardiac amyloidosis, and 116 with ATTR cardiac amyloidosis from 56 institutions [25]. Their three-stage ML cascade sequentially differentiated healthy volunteers from patients, HCM from cardiac amyloidosis, and AL from ATTR cardiac amyloidosis, with AUCs of 1.0, 0.99, and 0.92, respectively [25]. This study provides stronger evidence that clinical and CMR-derived imaging features can support subtype-aware classification, although prospective validation remains necessary.

CMR-based AI studies support CA detection, differentiation from phenotypically similar conditions, and ATTR-versus-AL subtype classification. However, subtype classification remains more difficult than binary CA detection because of overlapping imaging features between AL and ATTR. The current evidence is still limited by retrospective study designs, selected cohorts, and limited prospective external validation. Future work should focus on clinically representative multicenter cohorts and clearly defined subtype reference standards.

#### 4.2.3 Computed Tomography-Based Detection

Cardiac CT is an emerging source of opportunistic information in CA evaluation. CT is commonly obtained for adjacent indications, including transcatheter aortic valve replacement planning, coronary CT angiography, and evaluation of structural or ischemic heart disease. In these settings, CA may be suspected incidentally, creating an opportunity for opportunistic AI-based detection.

Lo Iacono et al. developed a radiomics-based ML pipeline using cardiac CT to differentiate CA from aortic stenosis in a cohort of 30 patients, including 15 with CA and 15 with aortic stenosis [26]. The model used myocardial radiomic features extracted from the LV wall and achieved accuracy, sensitivity, and specificity of 0.93 in this preliminary leave-one-out cross-validation setting. More recent CCTA radiomics work provided multicenter internal and external validation, achieving AUCs of 0.95 in training, 0.95 in internal testing, and 0.91 in external testing, outperforming myocardial CT attenuation alone [68]. Shiri et al. evaluated AI-based ATTR-CM detection in 263 patients with severe aortic stenosis using pre-TAVI data, including 27 confirmed ATTR-CM cases [69]. Among the evaluated modalities, 4D-CT strain had the strongest single-modality performance, with an AUC of 0.85, sensitivity of 0.90, and specificity of 0.74 [69].

CT-based AI is mainly attractive for opportunistic CA detection because CT is often obtained for valve assessment or coronary evaluation before CA is specifically suspected. This is particularly relevant in aortic stenosis and other hypertrophic phenotypes, where CA may coexist or be clinically overlooked, and CT-derived radiomic or functional features may contain information not captured by attenuation alone. However, compared with scintigraphy, echocardiography, and CMR, the CT evidence base remains earlier and more limited. Larger multicenter studies with clinically adjudicated CA labels, relevant mimics, subtype confirmation, and prospective validation are needed before CT-based AI can be considered more than diagnostic support for opportunistic case finding.

#### 4.2.4 Overall: Binary Detection Is Strongest for Scintigraphy, While Subtype-Aware Detection Remains Underdeveloped

Current detection evidence is strongest for narrowly defined imaging tasks with consistent labels, especially models applied to cardiac scintigraphy with bone-avid tracers that identify CA- or ATTR-CM-suggestive cardiac uptake in large and externally validated datasets. However, these models detect uptake patterns and should be interpreted as diagnostic support, not as replacements for monoclonal protein assessment, confirmatory imaging interpretation, and clinical adjudication. CMR and CT provide additional structural, functional, and tissue-level information, but their AI evidence remains more heterogeneous, often relying on smaller retrospective cohorts, selected comparator groups, and limited external validation. Subtype-aware detection is even less mature: ATTR-versus-AL classification is clinically important, but imaging-only models remain constrained by biological overlap and limited subtype-confirmed external datasets. Future studies should prioritize multimodal approaches that combine imaging, laboratory, and clinical data in cohorts with relevant mimics and patient-level adjudicated outcomes.

### 4.3 Quantification: Measuring Disease Burden

Once cardiac amyloidosis has been suspected or confirmed, the clinical question shifts from whether disease is present to how much disease is present, where it is concentrated, and whether it changes over time. This question has become more important as disease-modifying therapies make longitudinal assessment increasingly relevant. Visual Perugini grading remains part of established ATTR-CM image interpretation, whereas semiquantitative measures such as the heart-to-contralateral-lung ratio are better framed as planar uptake indices that may support disease-burden assessment and serial follow-up. However, these measures do not fully capture the volumetric extent of myocardial tracer burden or subtle longitudinal change. AI-enabled quantification has therefore advanced most clearly in nuclear imaging, where automated segmentation and standardized biomarker extraction can reduce dependence on manual region definition, acquisition-specific interpretation, and non-volumetric planar or semiquantitative measures.

#### 4.3.1 AI-Enabled SPECT/CT Quantification of Tracer Burden

The clearest AI-enabled quantification work comes from bone-avid tracer SPECT/CT. Miller et al. developed a deep learning workflow for 99mTc-PYP SPECT/CT in 299 patients undergoing imaging for suspected ATTR cardiac amyloidosis, including 83 confirmed ATTR-CA cases [30]. The model segmented cardiac chambers and myocardium from co-registered CT attenuation maps and applied these segmentations to attenuation-corrected SPECT images to quantify radiotracer activity. Rather than producing only a disease probability, the workflow extracted interpretable burden markers: target-to-background ratio, volume of involvement, and cardiac pyrophosphate activity, corresponding to relative uptake intensity, spatial extent of abnormal myocardial uptake, and composite uptake burden [30]. Among patients with ATTR-CA, 23 experienced cardiovascular death or heart-failure hospitalization, and the AI-extracted measures were associated with increased event risk after adjustment for age, with hazard ratios ranging from 1.41 to 1.84 per standard-deviation increase [30].

AI-based SPECT/CT quantification has also been extended to longitudinal assessment. Miller et al. evaluated serial 99mTc-PYP SPECT/CT quantification in 85 ATTR-CM patients with at least two scans [70]. In treated patients, volume of involvement decreased from a median of 100 to 51, and cardiac pyrophosphate activity decreased from 165 to 81, both with $P < 0.001$, over a median interval of 369 days [70]. Change in cardiac pyrophosphate activity showed a modest correlation with change in native T1 on CMR, with $\rho = 0.376$ and $P = 0.009$. After adjustment for age, treatment, and follow-up cardiac pyrophosphate activity, an increase in cardiac pyrophosphate activity during follow-up was associated with higher risk of cardiovascular death or heart-failure hospitalization, with an adjusted hazard ratio of 2.31 per standard-deviation increase [70]. These results support AI-enabled PYP SPECT/CT as a candidate tool for measuring longitudinal change in tracer burden, although the modest correlation with native T1 and retrospective design indicate that further validation is needed.

Spielvogel et al. evaluated AI-enabled 99mTc-DPD SPECT/CT quantification for treatment-response assessment in 45 ATTR-CM patients receiving disease-modifying therapy, including 37 with wild-type ATTR-CM and 8 with variant ATTR-CM [31]. The AI workflow extracted 26 thoracic SPECT/CT markers, including SUV metrics, retention index, amyloid-affected volume, and amyloid activity. After treatment, 17 of 26 markers, or 65%, significantly decreased, all with $P < 0.001$. Left ventricular SUVmax decreased from 18.6 to 14.1, and myocardial SUVmax decreased from 19.5 to 15.5 [31]. Several marker changes were associated with clinical improvement: reduced myocardial SUVpeak was associated with NT-proBNP reduction, $P = 0.007$, and reduced left ventricular amyloid activity was associated with NT-proBNP reduction, $P = 0.009$ [31]. Increased right ventricular amyloid-affected volume was associated with death or heart-failure hospitalization, with a hazard ratio of 3.19 and 95% confidence interval of 1.29–7.86 [31]. This study supports the ability of AI-enabled SPECT/CT to quantify treatment-associated changes in tracer burden, but the moderate cohort size means that these markers should be considered promising rather than established surrogate endpoints.

#### 4.3.2 AI-Enabled Cardiac Magnetic Resonance Imaging Quantification of T1 and Extracellular Volume

CMR provides a complementary approach to quantitative disease characterization through native T1 and extracellular volume (ECV), which reflect myocardial tissue properties and extracellular expansion rather than tracer uptake. AI can facilitate these measurements by automating myocardial segmentation and reducing the manual processing required for T1 and ECV analysis. Chang et al. evaluated a commercially available deep-learning algorithm for automated native T1 and ECV measurement in 95 participants with a range of cardiac conditions, including cardiac amyloidosis. Automated measurements showed close agreement with expert analysis, supporting the feasibility of AI-assisted quantitative CMR assessment [71].

More directly in CA, Hwang et al. evaluated automated ECV measurement in 300 patients undergoing CMR for differential diagnosis of left ventricular hypertrophy, including 50 patients with CA (39 AL and 11 ATTR). An ECV cutoff of 33.6% differentiated CA from other causes of LV hypertrophy with an accuracy of 85.6%. In patients with AL-CA, higher automated ECV was also associated with cardiovascular death or heart-failure hospitalization and improved risk stratification when added to the revised Mayo staging system [72]. These findings suggest that AI can make established CMR tissue-characterization measures more scalable and reproducible. However, native T1 and ECV reflect myocardial tissue alteration and extracellular expansion rather than amyloid protein identity and should therefore be interpreted as quantitative tissue-characterization markers rather than direct measures of amyloid deposition.

#### 4.3.3 AI-Enabled Functional Quantification from Echocardiography

Although nuclear imaging currently provides the most direct AI-based quantification of amyloid-related tracer burden, echocardiographic AI may support longitudinal monitoring of functional progression. Venneri et al. used a fully automated AI platform to analyze baseline and 12-month transthoracic echocardiograms in 752 patients with ATTR-CM [73]. The main AI-derived metric was left ventricular outflow tract velocity-time integral, a surrogate of stroke volume. During follow-up, a decrease in AI-derived LVOT-VTI of at least 5% over 12 months occurred in 377 patients, or 50.1%, and remained associated with all-cause mortality after adjustment for age, sex, TTR genotype, atrial fibrillation, NYHA class, and NAC stage, with a hazard ratio of 1.41 and 95% confidence interval of 1.13–1.76 [73]. Adding this threshold improved risk reclassification, with integrated discrimination improvement of 0.012 and continuous net reclassification improvement of 0.21, both $P < 0.001$ [73]. This study does not quantify amyloid deposition directly, but it shows that AI can reproducibly extract a functional marker of disease progression that is clinically meaningful in ATTR-CM.

#### 4.3.4 Overall Interpretation of Quantification Evidence

AI-based quantification in cardiac amyloidosis is strongest when the model output is an interpretable continuous biomarker rather than an opaque classification score. In nuclear imaging, AI can quantify tracer intensity, involved volume, and composite amyloid activity from SPECT/CT, with early evidence linking these measures to outcomes and treatment-associated change. In CMR, AI-assisted native T1 and ECV measurements provide scalable quantitative markers of myocardial tissue alteration and extracellular expansion, with emerging diagnostic and prognostic evidence. However, these measures reflect tissue characteristics rather than amyloid protein identity or direct amyloid burden. In echocardiography, AI-derived LVOT-VTI provides a scalable measure of functional progression rather than amyloid burden itself. Across modalities, these approaches remain primarily retrospective and require prospective validation, harmonized acquisition protocols, and standardized thresholds before they can be treated as validated monitoring endpoints. These results should therefore be interpreted as evidence of biomarker promise, not proof of clinical surrogacy. Before AI-derived SPECT/CT, PET, CMR, or echocardiographic measures can guide treatment escalation, switching, or trial endpoints, they require standardized acquisition protocols, multicenter repeatability data, longitudinal validation against clinical outcomes, and evidence that changes in the imaging marker reflect treatment-mediated clinical benefit.

### 4.4 Prognosis and Longitudinal Monitoring

Prognosis and longitudinal monitoring represent later-stage AI tasks in the CA pathway. Prognostic stratification estimates future risk after diagnosis, including death, heart-failure hospitalization, arrhythmia, transplant, or clinical deterioration. Longitudinal monitoring asks whether disease burden, function, or AI-derived risk signals change over time, particularly after treatment initiation. These tasks differ from detection because they require outcome linkage, serial assessment, calibration, and comparison with established biomarkers or staging systems.

Some diagnostic AI outputs also appear to encode prognostic information. Amadio et al. evaluated a previously developed diagnostic AI-ECG amyloid score in 2,533 patients with cardiac amyloidosis, including AL, wild-type ATTR, and hereditary ATTR disease [74]. The AI-ECG score remained prognostic on multivariable analysis, with a risk ratio of 2.0 in AL cardiac amyloidosis (95% CI, 1.58–2.55) and 2.7 in ATTR cardiac amyloidosis (95% CI, 1.81–4.24) [74]. Similarly, Spielvogel et al. reported that AI-positive cardiac uptake on multicenter cardiac scintigraphy with bone-avid tracers carried prognostic value in addition to diagnostic value [28]. These studies suggest that diagnostic AI scores may capture disease severity, but they should be interpreted as risk markers rather than treatment-decision tools.

Dedicated prognostic modeling remains limited but is clinically important. Wang et al. developed a fully automated CMR deep learning model for individualized prognosis prediction in light-chain cardiac amyloidosis using whole-heart LGE analysis [75]. In a cohort of 394 AL-CA patients, the model achieved a C-index of 0.91 and a 2.6-year survival AUC of 0.95 in the held-out test set, outperforming the Mayo staging model [76] (C-index 0.65; AUC 0.71) [75]. Zhou et al. developed a multicenter LGE-CMR radiomics model for all-cause mortality prediction in 120 CA patients from three institutions [77]. Together, these studies suggest that AI can extract prognostic information from LGE-CMR beyond conventional visual interpretation, although broader external validation, calibration assessment, and prospective clinical testing remain necessary.

Longitudinal monitoring evidence is most clinically meaningful when serial AI-derived change is linked to biomarkers, function, or outcomes. In nuclear imaging, AI-derived 99mTc-DPD and 99mTc-PYP SPECT/CT markers have been applied to serial ATTR-CM studies, showing treatment-associated changes in tracer-burden measures and associations with outcomes [31,70]. In echocardiography, Venneri et al. used a fully automated AI platform to analyze baseline and 12-month transthoracic echocardiograms in 752 patients with ATTR-CM [73]. A decrease in AI-derived LVOT-VTI of at least 5% over 12 months occurred in 377 patients and was associated with worse survival after adjustment for age, sex, TTR genotype, atrial fibrillation, NYHA class, and NAC stage, with a hazard ratio of 1.41 [73]. This study does not quantify amyloid deposition directly, but it supports AI-derived functional change as a scalable marker of disease progression.

Overall, prognosis and longitudinal monitoring remain less mature than detection and static quantification, but they are increasingly important as CA care moves toward earlier diagnosis, disease-modifying therapy, and serial response assessment. Current evidence suggests that AI-derived ECG, CMR, echocardiographic, and nuclear imaging markers can capture outcome-related or trajectory-related information. However, most studies remain retrospective or early-stage. Future work should train models directly on clinically meaningful endpoints, report calibration, compare against established staging systems and biomarkers, validate externally, and test whether AI-derived changes can guide treatment escalation or serve as trial endpoints.

## 5. From Model Performance to Clinical Translation

Section 4 reviewed machine learning applications in cardiac amyloidosis according to clinical task, including screening, detection, quantification, prognosis, and treatment-response monitoring. Across these tasks, many studies report strong discrimination in selected datasets. However, high model performance alone does not establish clinical readiness. Translation depends on whether models are developed on representative cohorts, evaluated with clinically meaningful metrics, interpreted in a way that supports physician trust, and validated under conditions that resemble real-world deployment. This section therefore examines the methodological patterns that determine whether CA AI models can move from promising retrospective experiments to reliable clinical decision support. Five cross-cutting issues are especially important: feature representation, cohort design, evaluation metrics, interpretability, and translational readiness.

### 5.1 Features Used by CA AI Models: Latent Signals, Radiomics, Clinical Variables, and Interpretable Biomarkers

AI models in cardiac amyloidosis differ not only by modality or architecture, but also by the type of information they use. Across reviewed studies, input features can be grouped into four broad categories: latent features learned directly from raw data, radiomic texture and intensity features, engineered clinical or imaging variables, and predefined quantitative biomarkers (Fig. 4). This distinction matters because feature representation influences both model performance and the extent to which model outputs can be connected to clinical reasoning.

The least directly interpretable category consists of latent features learned from raw ECGs, echocardiographic videos, CMR images, or scintigraphy images. These features are not manually defined before training; instead, deep learning models learn internal representations from waveform-level or pixel-level data [14,15,18,19,23,24,27,28,40,43,58]. Such models may detect subtle patterns beyond routine visual interpretation, but the final prediction is often difficult to translate into a specific clinical measurement because it depends on hidden model representations. Radiomics provides a more explicit but still complex feature strategy. Radiomic pipelines extract predefined descriptors [26,44]. These features may capture myocardial heterogeneity in mimic-rich settings such as HCM, hypertensive heart disease, aortic stenosis, or suspected CA. However, individual radiomic variables are often difficult to interpret biologically. Their stability also depends on segmentation quality, acquisition protocol, reconstruction method, and external validation. A third category includes engineered clinical and imaging variables, such as wall thickness, LV mass, LGE presence, strain, ECG-derived parameters, laboratory biomarkers, routine echocardiographic measurements, and preprocedural CT measurements [25,33,51,63,65]. These variables are easier to interpret because they correspond to familiar clinical concepts. Their limitation is that they may miss subtle waveform, texture, or regional disease patterns captured by latent or radiomic features.

The most directly interpretable features are predefined quantitative biomarkers, particularly in nuclear imaging. SUV-based measures, target-to-background ratio, volume of involvement, cardiac pyrophosphate volume, and cardiac pyrophosphate activity correspond to measurable concepts of uptake intensity, spatial extent, and composite tracer burden [21,30,46,48–50]. Recent AI-enabled SPECT/CT studies extend this approach by automating extraction of these burden measures and applying them to longitudinal assessment [31,70]. These outputs are clinically useful because they can be interpreted as continuous disease-burden measurements rather than opaque probability scores.

Overall, feature representation affects both what a model can learn and how easily its output can be used clinically. Latent deep learning features may improve pattern recognition; radiomics may capture myocardial heterogeneity; engineered clinical variables anchor predictions to familiar measurements; and predefined nuclear biomarkers provide the clearest link to measurable disease burden. Future CA AI studies should report which feature types drive prediction, whether those features are stable across sites and protocols, and whether they correspond to clinically meaningful disease biology.

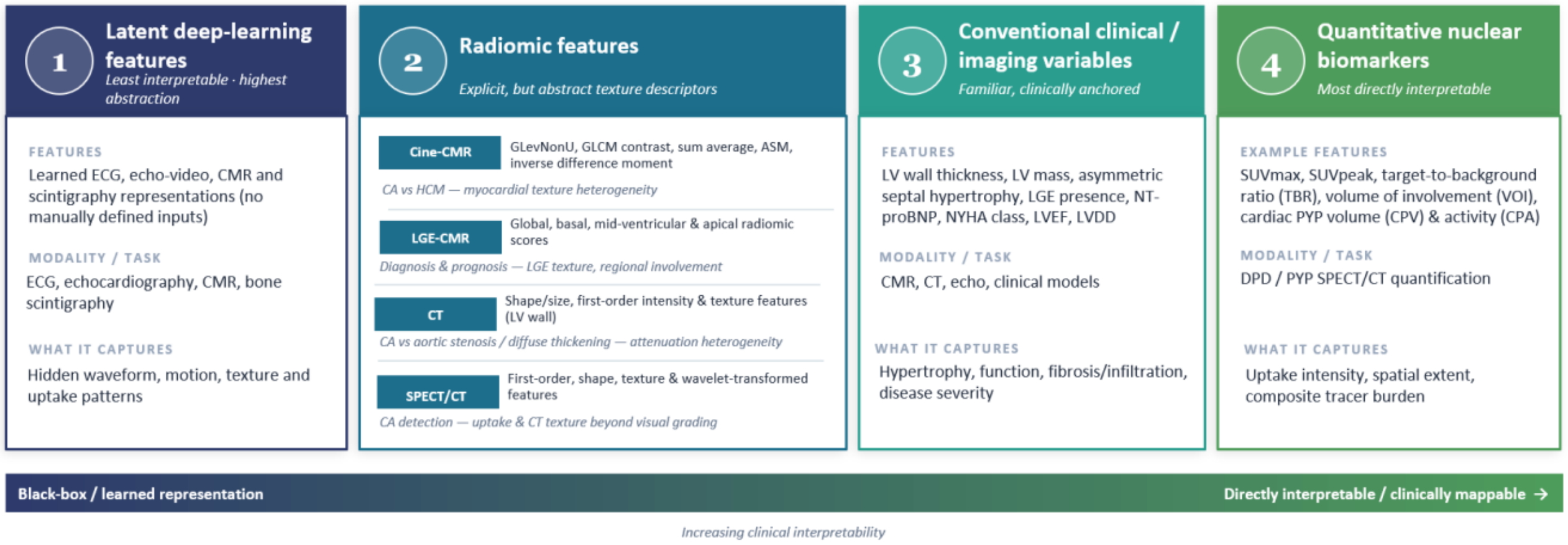

***Fig. 4*** *Feature representations in cardiac amyloidosis AI by clinical interpretability. The figure summarizes how model inputs range from latent deep-learning features to radiomics, conventional clinical variables, and quantitative nuclear biomarkers. Interpretability increases as features become more directly linked to measurable disease biology and tracer burden.*

## 5.2 Cohort Design and Generalizability

The apparent performance of CA AI models depends strongly on cohort design. Sample size, class balance, disease prevalence, comparator selection, scanner diversity, geography, and reference-standard quality all influence whether reported performance will generalize beyond the development dataset. Small proof-of-concept studies, such as CT radiomics, multimodal echo-EHR fusion, or AI-ECG treatment monitoring, are useful for hypothesis generation but cannot by themselves support strong claims about clinical readiness [26,32,78]. Comparator selection is especially important because CA often resembles more common cardiovascular disease. Models tested only against healthy controls or broad low-risk populations answer a different question from models tested against realistic mimics such as HFpEF, hypertensive LVH, aortic stenosis, HCM , and nonspecific heart failure [6,13,39]. ECG studies showing performance changes across different case and control definitions illustrate why comparator selection directly affects clinical meaning [40]. Generalizability also depends on whether models are evaluated across institutions, scanner vendors, tracers, reconstruction methods, acquisition protocols, disease subtypes, comparator diseases, and health-system settings. Larger and externally validated studies in ECG, echocardiography, and scintigraphy provide stronger evidence when they include independent sites, diverse acquisition conditions, and clinically relevant subgroups [14,18,19,25,28,34,41]. This is particularly important for nuclear imaging, where bone-avid tracers, scanners, reconstruction methods, and visual interpretation practices can vary across centers [9,10,28,46]. Future CA AI datasets should therefore be designed not only for size but also for clinical realism. Stronger datasets should include adequate numbers of AL and ATTR cases, relevant mimics, subtype-confirmed labels, patient-level separation between training and testing, and external validation across sites and acquisition protocols.

## 5.3 Evaluation of metrics and clinical utility

Evaluation in CA AI should move beyond AUC, sensitivity, and specificity. AUC measures discrimination, but it does not show whether a model is calibrated, useful at a clinical threshold, or effective in selecting patients for downstream testing. This is especially important for rare-disease screening, where positive predictive value depends strongly on prevalence. Studies should therefore report calibration, PPV, NPV, and performance at realistic deployment prevalence. The staged ECG-echocardiography workflow illustrates this point: ECG-only screening produced low positive predictive value in simulated surveillance, whereas a staged approach improved PPV by enriching the population before applying echocardiographic AI [14].

External validation and subgroup analysis should also be expected, because overall AUC can hide poor performance in clinically relevant groups, such as Hispanic patients, patients with LVH, or patients with left bundle branch block [22]. This matters because CA models are intended for heterogeneous real-world populations, not only curated development cohorts.

### 5.4 Interpretability, Reproducibility, Workflow Integration, and Regulatory Readiness

Clinical translation of CA AI requires more than high discrimination. Model outputs must also be interpretable, reproducible, and usable within real diagnostic workflows. Interpretability is commonly addressed through saliency maps, attention maps, feature-importance plots, or SHAP-style explanations [23,24]. These tools can show whether a model focuses on plausible cardiac regions or clinically relevant variables, but they do not necessarily explain the biological basis of prediction or define the next clinical action. Predefined quantitative biomarkers offer a more actionable form of interpretability because measures such as target-to-background ratio, volume of involvement, cardiac pyrophosphate activity, SUV metrics, and amyloid activity correspond to measurable concepts of tracer intensity, involved volume, and composite disease burden [30,31].

Reproducibility remains a parallel limitation. The current CA AI evidence base is stronger for model development than for implementation. Reproducibility requires clear reporting of inclusion criteria, case definitions, comparator groups, preprocessing, missing-data handling, train-test separation, threshold selection, calibration, external validation, subgroup analysis, and code or model availability when possible [79,80]. For imaging studies, acquisition and reconstruction details should be reported because differences in scanner vendor, sequence parameters, and reconstruction algorithms may affect learned features and quantitative biomarkers [9,42,46].
Workflow integration should be defined before deployment because screening, detection, quantification, prognosis, and treatment monitoring require different outputs, thresholds, validation cohorts, and safeguards. Regulatory readiness also remains early: broader cardiovascular AI experience shows that clearance does not guarantee clinical benefit or generalizability across settings [81]. Cardiac amyloidosis AI should therefore be evaluated with attention to intended use, clinical risk, performance monitoring, human oversight, and post-deployment drift. Trustworthy AI frameworks are useful because they emphasize fairness, universality, traceability, usability, robustness, and explainability across the AI lifecycle [82].

Human-in-the-loop design should be explicit for each clinical task. For screening, an AI alert should trigger structured review of red flags and appropriate confirmatory testing rather than automatic diagnosis. For image interpretation, AI output should assist readers by highlighting uptake localization, uncertainty, or quantitative burden, while final reports remain physician adjudicated. For prognosis and monitoring, AI-derived risk or change scores should be interpreted alongside NT-proBNP, troponin, functional status, treatment exposure, imaging findings, and patient preferences.

Multimodal foundation models, federated learning, and large language models remain future facing rather than established CA tools. Foundation models may eventually enable shared representations across ECG waveforms, echocardiographic videos, CMR, CT, SPECT/CT, PET, biomarkers, genomics, and clinical notes, but current CA evidence is not sufficient to support claims of clinical deployment. Federated learning could support privacy-preserving multi-institution training, especially for rare diseases such as AL-CA and ATTR-CM, but it does not by itself solve label quality, site bias, calibration, or subgroup fairness. Large language models may assist structured reporting, literature surveillance, trial matching, patient-facing education, and multidisciplinary documentation, but they should not be presented as autonomous diagnostic systems because of risks related to hallucination, citation reliability, privacy, and medicolegal accountability.

## 6. Roadmap for Clinical Translation

Across CA AI studies, models appear closest to clinical translation when the task is narrow, the imaging signal is clear, and the reference label is consistent, as seen with scintigraphy uptake detection and quantitative SPECT/CT biomarkers. In contrast, subtype classification, prognosis, and treatment-response monitoring remain less mature because they require stronger labels, outcome linkage, calibration, external validation, and workflow testing. The next phase of progress should therefore move beyond additional high-AUC retrospective classifiers toward clinically aligned multimodal, subtype-aware, and longitudinally validated systems that support suspicion, interpretation standardization, disease-burden quantification, and monitoring within established CA pathways.

### 6.1 Clinically Aligned Multimodal and Longitudinal Models

Future CA AI models should better reflect the multimodal structure of the diagnostic pathway. Single-modality models remain useful for focused tasks such as ECG screening, echocardiographic triage, scintigraphy uptake detection, and SPECT/CT quantification, but clinical decisions often require integration of imaging, laboratory testing, monoclonal protein assessment, clinical history, and, when needed, biopsy. Two implementation strategies should be compared directly. The first is staged risk enrichment, in which an accessible test identifies patients who should undergo more specific downstream testing. Goto et al. illustrated this approach by using ECG pre-screening before echocardiographic AI to improve positive predictive value in simulated deployment [14]. The second is direct multimodal fusion, where multiple data types are combined within one model. Feng et al. demonstrated the feasibility of combining echocardiographic videos with EHR-derived variables, although the small cohort keeps this as proof-of-concept evidence [32]. Future studies should test when staged workflows are preferable to direct fusion, rather than assuming that one design is always superior.

Subtype-aware modeling also remains a priority. Imaging-only ATTR-versus-AL classification is not ready to replace established diagnostic criteria, but models that combine imaging with monoclonal protein testing, genetic information when appropriate, and clinical adjudication may help reduce uncertainty and guide confirmatory testing [7,23,25]. Longitudinal modeling is equally important as disease-modifying therapies expand. Serial AI-ECG, AI-echocardiography, and AI-enabled SPECT/CT quantification should be evaluated as trajectories of disease progression, disease burden, and treatment-associated change, and compared with NT-proBNP, troponin, functional class, imaging measures, therapy status, and clinical outcomes [28,34,70,78].

### 6.2 Stronger Datasets, Labels, and Benchmarks

Clinical translation will depend on dataset quality as much as model architecture. Future CA AI studies need cohorts with realistic comparator groups, including HFpEF, hypertensive LVH, aortic stenosis, hypertrophic cardiomyopathy, suspected CA later excluded by confirmatory testing, and non-CA heart failure. Models tested only against healthy or broadly normal controls answer a different question from models tested in patients who create real diagnostic uncertainty. Stronger labels are especially important for subtype classification, prognosis, and treatment monitoring. Whenever possible, datasets should include monoclonal protein testing, clinically adjudicated ATTR diagnosis, biopsy confirmation, amyloid typing, genetic information when relevant, and outcome-linked endpoints [2,7,13,23,25,70,75,83]. This is necessary to avoid overclaiming from imaging-derived labels alone and to support patient-level diagnostic or prognostic conclusions. The field would also benefit from shared multicenter benchmarks, particularly for subtype-confirmed

CMR and SPECT/CT. Current studies are difficult to compare because they use different data sources, inclusion criteria, reference standards, tasks, and evaluation metrics. Shared definitions, harmonized endpoints, and prespecified external validation protocols would allow more meaningful comparison of architectures, radiomics methods, and quantitative biomarkers. Future cohorts should also capture demographic, clinical, and technical variation, including sex, age, race and ethnicity when available, amyloid subtype, comorbidities, scanner vendor, tracer type, reconstruction method, acquisition protocol, and health-system setting, so that subgroup performance, technical generalizability, and equity can be evaluated [22].

### 6.3 Evaluation Beyond AUC

Future studies should evaluate clinical usefulness, not only discrimination. AUC, sensitivity, and specificity remain necessary, but they are insufficient for deployment. Screening studies should also report calibration, positive predictive value, negative predictive value, and expected referral yield at realistic disease prevalence. This is especially important because CA is relatively uncommon in broad screening settings, where even a high-AUC model can generate many false positives if deployed without risk enrichment.

Threshold selection should be clinically justified and reported transparently. A screening model may require high sensitivity and high negative predictive value, whereas a confirmatory imaging-support model may require higher specificity. A prognostic model requires calibration and risk separation, while a treatment-monitoring marker requires reproducibility and sensitivity to change. Future studies should avoid tuning thresholds on test data and should report how thresholds would affect downstream testing, referral volume, false positives, and missed cases.

External validation should become a minimum expectation for clinically oriented models. Validation should be independent, preferably multicenter, and should test performance across institutions, scanners, vendors, tracers, acquisition protocols, and patient subgroups. Studies should also evaluate whether performance remains stable over time as referral patterns, imaging protocols, and disease prevalence change.

Subgroup analysis should be reported whenever sample size allows. At minimum, future studies should consider amyloid subtype, sex, age, race and ethnicity when available, LVH, conduction disease, heart-failure phenotype, aortic stenosis, scanner or vendor, tracer, and acquisition protocol. Studies that are too small for subgroup analysis should state this limitation explicitly rather than leaving fairness and generalizability unaddressed. Reporting and bias-assessment frameworks should be applied consistently [84,85].

### 6.4 Workflow Integration, Regulation, and Equity

Clinical translation will require prospective workflow evaluation, not only retrospective discrimination. Future studies should test whether AI improves diagnostic yield, time to diagnosis, time to treatment, unnecessary downstream testing, inter-reader variability, referral appropriateness, patient outcomes, or cost-effectiveness. Prospective silent trials and pragmatic implementation studies are especially important because they can evaluate real-world calibration, subgroup performance, alert burden, and clinician response before or during deployment. Recent implementation and prospective-evaluation efforts, including an AI-augmented ATTR-CM detection program and ongoing multicenter AI-toolkit or echocardiography studies, illustrate this shift toward workflow-level evaluation, although they should be interpreted as implementation frameworks until outcome results are available [60,61,86].

Regulatory planning should begin early because screening aids, image-interpretation tools, subtype-support models, prognostic models, and monitoring biomarkers have different intended uses and evidentiary requirements. Publications should clearly define the target population, clinical setting, intended user, model output, action threshold, expected clinician response, and potential harms; broader cardiovascular AI experience shows that clearance alone does not guarantee generalizability or clinical benefit [81]. Equity and post-deployment monitoring should also be built into implementation, since performance may vary across populations, health-system settings, scanners, protocols, and referral patterns. Deployed CA AI systems should therefore include performance surveillance, recalibration plans, human oversight, and subgroup monitoring, consistent with trustworthy-AI principles such as fairness, universality, traceability, usability, robustness, and explainability [82].

Global implementation requires separate equity lens. AI systems trained in high-income tertiary amyloidosis centers may not generalize to regions where access to CMR, SPECT/CT, PET tracers, amyloidosis specialists, technologists, standardized acquisition protocols, or confirmatory laboratory testing is limited. In low- and middle-income countries, the most useful AI tools may be those embedded in widely available modalities such as ECG, basic echocardiography, and structured clinical data, provided that downstream referral pathways, monoclonal protein testing, and confirmatory imaging are feasible. Future studies should therefore evaluate not only accuracy but also infrastructure requirements, workforce needs, cost, data representativeness, model maintenance, and the risk that AI could widen diagnostic inequities if deployed only where advanced imaging and specialist services are already available.

## 7. Conclusion

In conclusion, AI has shown meaningful potential across the cardiac amyloidosis pathway, particularly for suspicion-raising screening, automated uptake detection, and interpretable disease-burden quantification. However, subtype-aware classification, prognostic risk prediction, and treatment-response monitoring remain less mature and require stronger labels, outcome linkage, calibration, and external validation. Future progress will depend on larger multicenter datasets, realistic comparator groups, prospective workflow evaluation, and integration with established diagnostic pathways so that AI supports, rather than replaces, expert clinical decision-making. Overall AI in cardiac amyloidosis is moving from isolated detection models toward clinically aligned, multimodal, interpretable, and longitudinal systems. The strongest near-term value is likely to come from risk enrichment, standardized imaging interpretation, reproducible quantitative biomarker extraction, and workflow support. The field should avoid equating retrospective high AUC with clinical utility and should prioritize prospective evaluation, external validation, subgroup fairness, regulatory-grade documentation, and evidence that AI changes clinical decisions or outcomes.

## Funding

This research was supported in part by grants from the National Institutes of Health (1R15HL172198 and 1R15HL173852).

## Competing interests

Dr. Weihua Zhou reports consulting fees from Pressiant Health Inc. and SynterMed Inc., outside the submitted work. All other authors declare no competing interests.

# Supplementary Tables

## 1. Studies in cardiac amyloidosis, organized by clinical task

### A. Screening and risk enrichment studies in cardiac amyloidosis

| Study | Modality / input | Cohort / design | Model / approach | Main reported result | Validation level | Target / label | Contribution |
|---|---|---|---|---|---|---|---|
| Schrutka et al. (2021) | ECGI-derived electrical maps and **surface** ECG | Exploratory Vienna single-center cohort of CA patients and controls | Machine learning used to translate electroanatomical mapping patterns into surface ECG features | Cardiologist recognition of CA improved after training on ML-derived ECG patterns; reported AUC improved from 0.69 before training to 0.97 after training | No external validation | Mixed CA, including AL and ATTR | Mechanistic early ECG work showing that CA has recognizable electrical signatures; best framed as hypothesis-generating rather than as a deployable AI-ECG screening model |
| Grogan et al. (2021) | 12-lead ECG, single-lead V5, and 6-lead ECG subsets | 2,541 CA patients and 2,454 age- and sex-matched controls from Mayo Clinic | Deep neural network AI-ECG classifier | AUC 0.91 for 12-lead ECG, 0.86 for single-lead V5, and 0.90 for 6-lead ECG; identified 59% of prediagnosis cases more than 6 months before clinical diagnosis | Internal train/validation/holdout testing | AL and ATTR CA | Established that routine ECG contains a detectable CA signal and may flag some patients before clinical diagnosis, but evidence remains single-system and case-control enriched |
| Harmon et al. (2023) | 12-lead ECG | 440 newly diagnosed CA patients and 6,600 age- and sex-matched controls | Post-development validation of an AI-ECG amyloidosis model | Overall AUC 0.84; lower performance reported in Hispanic patients and in patients with LVH or LBBB | Single-institution post-development validation | AL and ATTR CA | Important generalizability study showing that acceptable overall AI-ECG performance can hide clinically relevant subgroup weaknesses |
| Haimovich et al. (2023) | 12-lead ECG; single-lead ECG models also evaluated | 50,709 patients with LVH-associated diseases, including 304 CA, 1,056 HCM, 20,802 hypertension, and 446 aortic stenosis cases | Pretrained CNN-derived ECG representations with logistic-regression classification (LVH-Net) | CA AUROC **0.95**; HCM 0.92; aortic stenosis 0.90; hypertensive LVH 0.76 | Retrospective validation in a large multi-institutional health-system cohort | CA among clinically relevant LVH etiologies | Important mimic-rich ECG study showing that AI can distinguish CA within a heterogeneous hypertrophic-disease population rather than only from normal controls |
| Goto et al. (2021) | ECG and echocardiographic video | ECG cohorts from 3 academic centers and echocardiography cohorts from 5 centers in the US and Japan | Separate CNN models for ECG and echocardiographic video; simulated staged ECG-to-echo workflow | ECG C-statistic 0.85–0.91; echo C-statistic 0.89–1.00; simulated ECG pre-screening improved echo PPV from about 33% to 74–77% at 67% recall | Multi-institution retrospective validation | Mixed AL and ATTR CA | Major multicenter risk-enrichment study showing that staged screening is more clinically plausible than isolated single-test classification |
| Vrudhula et al. (2024) | ECG waveforms | Approximately 1.3 million ECGs from 341,989 patients | Methodological analysis of AI-ECG case and control definitions | Matched held-out AUC varied from 0.660 to 0.898; broader real-world testing showed performance drops as low as 0.467 depending on training and testing definitions | Real-world population testing | Not subtype-specific CA screening | Critical cautionary study showing that AI-ECG performance can be strongly inflated or degraded by control selection, cohort enrichment, and deployment population |
| González-López et al. (2026) | Standard 12-lead ECG | 20,754 quality-controlled ECGs from 2,901 individuals, including 585 ATTR-CM patients and 2,316 controls | Willem AI ECG platform | AUC 0.88; sensitivity 80.7%; specificity 78.5%; similar performance in ATTRwt and ATTRv subgroups | Retrospective validation; no prospective workflow validation reported | ATTR-CM, including ATTRwt and ATTRv | Extends AI-ECG screening specifically to ATTR-CM and relevant subgroups, but prospective clinical utility remains unproven |
| Chang et al. (2024) | Routine echocardiographic measurements | 3,603 echocardiogram studies from 636 patients | Random forest using 19 routinely reported echocardiographic variables | AUC 0.84; sensitivity 0.82; specificity 0.73; PPV 0.76 | Internal testing only | Mixed CA | Demonstrates that structured echo measurements can support CA suspicion without end-to-end video DL, but repeated studies per patient and absent external validation limit clinical certainty |
| Duffy et al. (2025) | PLAX and A4C echocardiographic videos | 574 CA patients and 979 controls across international sites | EchoNet-LVH computer vision model | AUC 0.896; sensitivity 0.644, specificity 0.988, and PPV 0.968 at a prespecified high-specificity threshold | International multisite retrospective validation | Mixed CA | Strong externally validated echo AI study designed for high-specificity diagnostic enrichment rather than replacement of confirmatory testing |
| Ioannou et al. (2026) | Echocardiography with AI-derived measurements and video DL | 5,776 patients, including 2,756 CA and 3,020 controls across UK, Taiwan, US, and Japan cohorts | AI-derived multiparametric echo score and fully automated deep-learning video model | AI-derived score accuracy about 79.5–79.7% in external cohorts; DL video model accuracy 96.2% in internal validation, 95.8% in internal testing, and about 87.5–88.4% in external cohorts | Large international multicenter validation | Mixed CA | Large international study showing the maturity gradient within echo AI: interpretable measurement-based scores are useful, but end-to-end video DL achieved higher diagnostic performance |
| Slivnick et al. (2025) | Single apical four-chamber echocardiographic video clip | 597 CA patients and 2,122 controls, with external validation across 18 sites | Deep learning model using one routine A4C clip | AUROC about 0.93; sensitivity about 85%; specificity about 93% after excluding uncertain outputs | External validation across 18 sites; multivendor and multiethnic evaluation | Mixed AL and ATTR CA | Operationally attractive screening model because it uses one routinely acquired echo clip; supports scalable triage but still requires confirmatory diagnostic testing |
| Agibetov et al. (2020) | Routine laboratory parameters | CA-related heart-failure and non-CA heart-failure cohorts | Machine learning using routine laboratory variables | Demonstrated proof-of-concept discrimination of CA-related heart failure using standard laboratory parameters | Internal validation only | CA-related HF | Early non-imaging screening study suggesting that routine laboratory data may contain CA-associated systemic signals; useful as proof-of-concept rather than clinical-ready evidence |
| Pan et al. (2024) | Routine blood tests and clinical variables | 6,563 patients with LVH, including 261 CA cases | XGBoost, random forest, and logistic regression comparison | XGBoost AUC 0.95; sensitivity 0.92; specificity 0.95; F1 score 0.89 | Internal train/test validation only | Mixed CA in an LVH population | Supports non-imaging CA case finding using routine clinical data, but should be interpreted as internally validated evidence requiring external and prospective confirmation |
| Huda et al. (2021) | Claims and EHR-derived diagnosis-code patterns | 1,071 ATTRwt-CM cases and 1,071 HF controls for derivation, with additional claims and health-system validation cohorts | Random forest using administrative and diagnosis-code features | AUROC 0.93 in the derivation test set and 0.80 in a broader real-world HF cohort | Multiple retrospective validation cohorts, including health-system EHR testing | ATTRwt-CM / ATTR-CM risk | Shows that routinely collected health-system data can identify patients at risk for ATTRwt-CM, while also illustrating performance decline outside curated development settings |
| Mitchell et al. (2022) | EHR-derived clinical and administrative variables | Academic medical-center population screened for possible ATTR-CM | Machine-learning-adapted clinical algorithm | Identified patients with possible ATTR-CM for further clinical evaluation | Single-center implementation-oriented retrospective study | Possible ATTR-CM | Demonstrates feasibility of applying EHR-based risk enrichment within an academic medical center, but does not establish definitive clinical outcome benefit |
| Peters et al. (2023) | Hospitalized heart-failure registry and EHR variables | GWTG-HF hospitalized HF population | Adapted wild-type ATTR-CM machine-learning model | Evaluated ATTR-CM risk enrichment among hospitalized HF patients | Registry-based validation / implementation study | ATTRwt-CM risk in HF | Extends EHR-based screening into hospitalized HF populations and highlights the need to evaluate model yield in realistic care settings |
| Oikonomou et al. (2025) | Longitudinal ECG and echocardiography | 984 YNHHS and 806 Houston Methodist patients referred for nuclear amyloid testing; 7,352 TTEs and 32,205 ECGs | AI-ECG and AI-echo probability trajectories before nuclear testing | AI probabilities diverged up to 3 years before ATTR-CM testing; double-negative sensitivity 90.9% and 85.7%; double-positive specificity 85.5% and 88.9% across cohorts | Two-health-system retrospective longitudinal validation | ATTR-CM | Extends screening from one-time classification to longitudinal risk enrichment before confirmatory nuclear testing; useful bridge between screening and disease-trajectory modeling |
| Feng, Sivak and Krishnamurthy (2024) | PLAX/A4C echocardiographic videos plus EHR variables | 41 patients | Transformer-based multimodal fusion model | AUROC 0.94 in 5-fold cross-validation | No external validation | Mixed CA | Small proof-of-concept for echo-video and EHR fusion; useful for feasibility but too small to support claims of mature clinical screening |
| Jain et al. (2026) | ECG waveforms, echocardiographic measurements, demographics, and diagnosis-code patterns | Retrospective development and external testing plus prospective single-arm workflow | ATTRACTnet multimodal AI screening program | AUROC 0.85 internally and 0.82 externally; in the prospective workflow, 24 of 50 AI-flagged tested patients were diagnosed with ATTR-CM and 21 started treatment | External testing plus prospective non-randomized single-arm clinical workflow | ATTR-CM | Most implementation-facing screening evidence in this section; demonstrates diagnostic yield after AI-augmented referral, but randomized outcome evidence is still absent |
| TRACE-A / UCSF (ongoing) | Routine ECG, point-of-care ultrasound, and transthoracic echocardiography | Ongoing multicenter observational study across diverse US health systems | AI algorithms applied to routine cardiovascular testing to estimate underdiagnosis of ATTR-CM | Peer-reviewed outcome results not yet available | Ongoing observational deployment study | ATTR-CM underdiagnosis | Important future-direction study for real-world deployment; should be cited as ongoing evidence only, not as completed validation |

## B. Detection studies in patients with suspected cardiac amyloidosis

| Study | Modality | Cohort / sample | Model / approach | Main reported result | Validation | Target / subtype / label | Contribution |
|---|---|---|---|---|---|---|---|
| Halme et al. (2022) | 2D whole-body 99mTc-HMDP bone scintigraphy | 1,334 patients; 47 had Perugini grade ≥2 cardiac uptake | Two custom CNN architectures compared with VGG16, ResNet50, InceptionV3, and MobileNet; automated Perugini grading | AUC ≥0.85 for four-class Perugini grading; AUC 0.88–0.90 for grade <2 vs ≥2 uptake detection; AUC 0.94 for grade 3 vs grades 0–2 | Cross-validation; no independent external cohort | Cardiac uptake / Perugini grade suggestive of ATTR | Early AI scintigraphy study demonstrating automated detection and grading of cardiac uptake; supports interpretation standardization but does not replace patient-level ATTR diagnosis |
| Eckstein et al. (2022) | Non-contrast CMR-derived biatrial and right-ventricular strain/function | 107 subjects: 43 CA, 20 HCM, 44 healthy controls | 41 CMR-derived functional/strain features evaluated with DT, KNN, linear SVM, and RBF-SVM; PCA also assessed | Best RBF-SVM: 90.9% accuracy, AUC 0.996, sensitivity 100%, precision 94% | Retrospective single-center study; internal supervised evaluation only | CA vs HCM vs healthy controls | Shows that engineered multi-chamber functional and strain features can support CA detection without contrast-based tissue characterization; promising but limited by small single-center cohort |
| Delbarre et al. (2023) | Whole-body 99mTc bone scintigraphy | 3,048 images for development/internal testing and 1,633 images for external testing | CNN using image-level labels; Perugini grade ≥2 defined as positive | Sensitivity 98.9% in cross-validation and 96.1% in external validation; specificity 99.5% in both; AUC 0.999 in both settings | External validation | Significant cardiac uptake suggestive of CA / ATTR-CM workup | Strong whole-body scintigraphy AI study showing that clinically relevant cardiac uptake can be detected without manual myocardial segmentation; supports detection and standardization, not replacement of the full diagnostic algorithm |
| Spielvogel et al. (2024) | Bone scintigraphy using 99mTc-PYP, 99mTc- | 16,241 patients and 19,401 scans from 9 centers in 4 countries | CNN-based cardiac uptake detection pipeline with thorax cropping | AUC 1.000 in Austrian cross-validation, 0.997 in the United Kingdom, 0.925 in China, and 1.000 in | International, multicenter, cross-country, cross-tracer validation | Cardiac uptake suggestive of CA, most relevant to | Strongest nuclear detection evidence in this section because of scale, cross-tracer validation, and multicenter |
| | DPD, and 99mTc-HMDP | | | Italy; AI AUC 0.997 versus physician AUC 0.946 in reader comparison | | ATTR-CM workup | design; also reported mortality association, but this should be interpreted as secondary prognostic association |
| Salimi et al. (2025) | Total-body scintigraphy | Six datasets from 12 cameras, including MDP, DPD, HDP, and PYP datasets; external datasets from multiple centers | Fully automated DL pipeline for localization, ATTR-CM detection, and Perugini-based scoring | Internal detection/scoring models achieved AUC >0.95 and F1 >0.90; external detection AUCs 0.93, 0.95, and 1.00; external scoring AUCs 0.95, 0.83, and 0.96 | Multicenter, multiscanner, multitracer evaluation | ATTR-CM detection and scoring | Extends scintigraphy AI beyond binary uptake detection toward automated localization and scoring; useful for standardizing nuclear image interpretation across heterogeneous acquisition settings |
| Mo et al. (2025) | SPECT/CT using PYP or DPD | 290 patients in a multicenter retrospective cohort | Radiomics with reproducibility filtering, LASSO feature selection, and stacking ensemble using SVM, random forest, gradient-boosted decision tree, and AdaBoost | AUC 0.871 in internal validation, 0.824 in external test cohort 1, and 0.839 in external test cohort 2 | Two external test cohorts; preprint at present | Mixed CA / SPECT-CT diagnostic classification | Emerging SPECT/CT radiomics evidence suggesting added diagnostic value beyond conventional visual or semiquantitative interpretation; should be clearly marked as preprint or emerging evidence until peer-reviewed |
| Martini et al. (2020) | CMR with late gadolinium enhancement | 206 subjects with suspected CA | Deep learning on LGE images; comparator ML using structured CMR features | DL AUC 0.982; structured-feature ML AUC 0.952; no statistically significant difference between approaches | No external validation | Mixed CA | Early CMR DL detection study showing strong performance, while also showing that structured CMR features can perform similarly in a moderate-sized cohort |
| Agibetov et al. (2021) | CMR including LGE, MOLLI, and cine protocols | 502 patients, including 82 CA cases | CNN-based fully automated CMR diagnosis | Best fine-tuned CNN AUC 0.96; sensitivity 94%; specificity 90% | 10-fold cross-validation only | Mixed CA | Demonstrated feasibility of fully automated multiprotocol CMR-based CA diagnosis, but evidence remains cross-validated rather than externally validated |
| Zhou et al. (2022) | LGE-CMR radiomics | Retrospective multicohort diagnostic study | Radiomics feature extraction with ML classification | Radiomics provided incremental diagnostic information beyond visual assessment and conventional quantitative CMR parameters; radiomics scores correlated with CA severity | Multicohort retrospective validation | Mixed CA | Supports CMR radiomics as diagnostic support and disease-pattern characterization, but not as a stand-alone diagnostic replacement |
| Jiang et al. (2022) | Non-contrast cine-CMR | CA and HCM comparator cohorts | Cine-CMR radiomics with ML classification | Combined radiomics plus conventional MR model improved validation AUC to 0.89 compared with 0.79 for conventional MR metrics alone | Internal validation | CA versus HCM mimic setting | Clinically relevant because HCM is a major phenotypic mimic of CA; supports the value of mimic-rich evaluation rather than CA-versus-normal classification |
| Cockrum et al. (2024) | CMR with cine and LGE | Large Cleveland Clinic CMR cohort with internal validation and external CA testing | Vision transformer deep learning model | Reported high diagnostic discrimination for CA versus other cardiomyopathy referrals, including strong performance in equivocal CMR cases | Internal validation plus external CA cohort testing | Mixed CA versus other myocardial disease referrals | Expands CMR detection evidence beyond CNNs to attention-based vision transformer modeling; useful as possible second-opinion support, but not proof that transformers are clinically superior to simpler models |
| Germain et al. (2023) | CMR with cine and LGE | 120 patients, including 70 AL and 50 ATTR | VGG16 CNN models for cine and LGE images | Cine-CNN accuracy 0.750 and AUC 0.839; LGE-CNN accuracy 0.611 and AUC 0.679; best human reader AUC 0.714 | 5-fold cross-validation only; no external validation | AL versus ATTR subtype-aware detection | Important first DL CMR study directly addressing AL-versus-ATTR classification; performance was not sufficient for stand-alone clinical use, supporting the abstract's claim that subtype-aware classification remains early |
| Weberling et al. (2025) | Multiparametric multivendor CMR | 400 participants from 56 institutions | Three-stage ML cascade using semiautomated CMR data and patient demographics | Cascade AUCs of 1.00 for healthy versus non-healthy, 0.99 for HCM versus CA, and 0.92 for AL versus ATTR | Multicenter and multivendor retrospective validation | Healthy controls, HCM, AL-CA, ATTR-CA | Major subtype-aware CMR ML study using a clinically relevant cascade that includes HCM as a key mimic; promising but still retrospective and should be interpreted as diagnostic support rather than treatment-directing classification |
| Lo Iacono et al. (2023) | Cardiac CT | 30 patients, including 15 CA and 15 aortic stenosis | CT radiomics with ML classification | Preliminary leave-one-out performance around 0.93 for accuracy, sensitivity, and specificity | No external validation | Mixed CA versus aortic stenosis | Early CT radiomics proof-of-concept; clinically interesting because CA and aortic stenosis overlap, but limited by very small sample size |
| Meng et al. (2025) | Coronary CT angiography | Multicenter cohort from five hospitals of patients with diffuse myocardial thickening | CT radiomics model using myocardial radiomic features | AUC 0.95 in training, 0.95 in internal testing, and 0.91 in external testing; outperformed myocardial CT attenuation alone | External cohort from four hospitals | Mixed CA among hypertrophic phenotype patients | Stronger CT evidence than earlier proof-of-concept work; supports opportunistic diagnostic support in hypertrophic-phenotype CCTA populations |
| Shiri et al. (2025) | Pre-TAVI data including 4D cardiac CT strain, CT radiomics, ECG, echocardiography, and clinical variables | 263 patients with severe aortic stenosis, including 27 confirmed ATTR-CM cases | Multimodality AI models using routinely collected preprocedural data | 4D cardiac CT strain showed the strongest single-modality performance, with AUC about 0.85 | Single-center prospective cohort | ATTR-CM in severe aortic stenosis | Supports opportunistic ATTR-CM detection in valve-evaluation populations; should not be framed as a general CT screening model |

## C. Quantification of disease burden

| **Study** | **Modality** | **Cohort / sample** | **Model / approach** | **Main reported result** | **Validation** | **Target / subtype / label** | **Contribution** |
|---|---|---|---|---|---|---|---|
| Caobelli et al. (2019) | 99mTc-DPD SPECT/CT | Patients with suspected ATTR cardiac amyloidosis | Conventional quantitative SPECT/CT; not AI | Quantitative 99mTc-DPD SPECT/CT was feasible and enabled objective assessment of myocardial uptake; SUV-based measures correlated with Perugini visual score | No external validation | ATTR-relevant myocardial uptake | Foundational non-AI quantitative SPECT/CT study showing that objective uptake measures can complement visual grading and support later AI-based biomarker extraction |
| Ramsay and Cuscaden (2019) | Quantitative SPECT/CT | Technical / perspective article | Non-AI technical framework | Not a model-performance study | Not applicable | ATTR-relevant quantitative imaging | Useful background article articulating the role of quantitative SPECT/CT for disease-burden assessment; cite as rationale, not as primary validation evidence |
| Watanabe et al. (2021) | 99mTc-PYP SPECT/CT | Single-center ATTR-CA cohort | Conventional volumetric quantification; not AI | Cardiac pyrophosphate volume and cardiac pyrophosphate activity correlated with LVEF, posterior wall thickness, and QRS duration | No external validation | ATTR-CA | Defined volumetric PYP biomarkers that later AI workflows sought to extract more reproducibly and at scale |
| Ayers, Peruri and Bourque (2021) | Quantitative SPECT | Editorial / clinical rationale | Non-AI conceptual framework | Not a model-performance study | Not applicable | ATTR-CM | Emphasized the potential role of quantitative SPECT for diagnosis, prognosis, and monitoring of disease progression; useful as framing rather than primary evidence |
| Chang et al. (2022) | CMR native T1 mapping and extracellular volume (ECV) | 95 participants with a range of cardiac conditions, including cardiac amyloidosis | Commercially available deep-learning algorithm for automated myocardial segmentation and measurement of native T1 and ECV | Automated measurements showed close agreement with expert analysis for both native T1 and ECV | Retrospective technical validation; not a dedicated CA-only cohort | Quantitative CMR measurement automation; included amyloidosis cases | Enabling study showing that AI can automate native T1 and ECV measurement with strong agreement to expert analysis, supporting more scalable and reproducible quantitative CMR assessment |
| Hwang et al. (2025) | CMR extracellular volume (ECV) | 300 patients undergoing CMR for differential diagnosis of left ventricular hypertrophy, including 50 patients with CA (39 AL and 11 ATTR) | Automated ECV measurement using AI-assisted CMR analysis | ECV cutoff of 33.6% differentiated CA from other causes of LV hypertrophy with 85.6% accuracy; in AL-CA, higher ECV was associated with cardiovascular death or heart-failure hospitalization and improved risk stratification when added to the revised Mayo staging system | Retrospective clinical cohort | CA vs other LVH causes; prognostic subgroup analysis in AL-CA | Most clinically relevant CMR-quantification study in this section, showing that AI-assisted ECV can support both diagnostic differentiation and prognostic stratification, while remaining a tissue-characterization marker rather than a direct measure of amyloid deposition |
| Bhattaru et al. (2024) | Bone scintigraphy SPECT/CT using PYP or HDP | 77 primary-cohort subjects and 93 external-validation subjects | Deep learning segmentation and automated uptake classification | Automated segmentation showed high agreement with ground-truth myocardial segmentation and supported automated identification of myocardial uptake suggestive of ATTR-CM | External validation cohort | ATTR-CM-suggestive myocardial uptake | Provides the segmentation foundation needed for reproducible SPECT/CT quantification and reduces dependence on manual myocardial region definition |
| Miller et al. (2024) | 99mTc-PYP SPECT/CT | 299 patients undergoing PYP SPECT/CT for suspected ATTR-CA, including 83 confirmed ATTR-CA cases | Deep learning CT attenuation-map segmentation followed by quantitative SPECT biomarker extraction | CPA AUC 0.989, VOI AUC 0.988, and TBR AUC 0.979 for ATTR-CA diagnosis; all three biomarkers were associated with cardiovascular death or heart-failure hospitalization after age adjustment | No external validation | ATTR-CA | Flagship AI-enabled quantification study producing interpretable biomarkers of uptake intensity, involved volume, and composite tracer burden; central evidence for the abstract's claim that SPECT/CT quantification is clinically translatable |
| Miller et al. (2026) | Serial 99mTc-PYP SPECT/CT | Patients with cardiac amyloidosis who underwent at least two PYP studies | AI-driven longitudinal PYP quantification using deep learning-derived segmentation and | Serial CPA and VOI tracked change over time and were evaluated against multimodality imaging parameters and outcomes; | Retrospective longitudinal validation | ATTR-CM / cardiac amyloidosis with serial PYP imaging | Extends AI-driven PYP quantification from single-time-point disease-burden measurement to longitudinal disease tracking and provides a bridge toward |
| | | | biomarker extraction | increasing CPA was associated with worse outcome | | | treatment-response monitoring |

## D. Prognostic stratification and risk prediction

| Study | Modality | Cohort / sample | Model / approach | Main reported result | Validation | Target / subtype / label | Contribution |
|---|---|---|---|---|---|---|---|
| Zhou et al. (2023) | LGE-CMR radiomics | 120 CA patients from 3 institutions; 40 deaths during median follow-up of 12.9 months | Radiomic feature extraction from global and segmental LV myocardium with ML-based risk modeling | Radiomic risk score improved all-cause mortality prediction beyond clinical staging and separated survival groups on Kaplan-Meier analysis | Multicenter retrospective cohort; no prospective validation | Mixed CA; all-cause mortality prediction | Dedicated CMR radiomics prognosis study showing that LGE-derived texture and intensity patterns contain risk information beyond conventional clinical staging |
| Wang et al. (2025) | CMR with LGE | 394 AL-CA patients from 3 hospitals; 315 training and 79 testing patients | Fully automated whole-heart DL analysis using automated segmentation, contrastive pretraining, and survival prediction from LGE images | Testing C-index 0.91 and AUC 0.95 for 2.6-year survival prediction; outperformed Mayo staging, which had testing C-index 0.65 and AUC 0.71 | Retrospective multicenter development/testing; no prospective clinical validation | AL-CA; individualized overall-survival prediction | Strongest dedicated DL prognosis study in CA; supports the abstract's claim that prognostic AI is promising but still early because prospective utility against established staging systems remains unproven |
| Amadio et al. (2025) | AI-enhanced 12-lead ECG | 2,533 CA patients, including 1,834 AL, 530 ATTRwt, and 169 ATTRv patients | Previously developed diagnostic AI-ECG amyloid score evaluated as a survival risk marker | Higher A2E score was associated with increased all-cause mortality; adjusted risk ratio 2.0 in AL and 2.7 in ATTR groups | Retrospective prognostic evaluation in established CA | AL and ATTR CA; all-cause mortality | Shows that diagnostic AI-ECG outputs can encode severity-related prognostic information, but should be interpreted as risk markers rather than dedicated treatment-decision models |
| Miller et al. (2024) | 99mTc-PYP SPECT/CT | 299 patients undergoing PYP SPECT/CT for suspected ATTR-CA; 83 confirmed ATTR-CA cases | Deep learning CT segmentation followed by quantitative SPECT biomarker extraction | TBR, VOI, and CPA were associated with cardiovascular death or heart-failure hospitalization after age adjustment; HRs ranged from 1.41 to 1.84 per SD increase | Single-center retrospective cohort; no external validation | ATTR-CA; secondary outcome association | Secondary prognostic evidence from an AI quantification study showing that interpretable tracer-burden biomarkers carry outcome signal |
| Spielvogel et al. (2024) | Bone scintigraphy using PYP, DPD, and HMDP | 16,241 patients and 19,401 scans from 9 centers in 4 countries | AI scintigraphy detection model for CA-suggestive cardiac uptake | AI-positive uptake status was associated with mortality across cohorts | International multicenter, cross-country, cross-tracer validation | CA-suggestive uptake, most relevant to ATTR-CM workup | Secondary prognostic evidence from a large detection study; useful for showing outcome association, but not a dedicated prognosis model |

## E. Treatment-response monitoring and longitudinal disease tracking

| Study | Modality | Cohort / sample | Model / approach | Main reported result | Validation | Target / subtype / label | Contribution |
|---|---|---|---|---|---|---|---|
| Schlesinger et al. (2025) | Serial 12-lead AI-ECG | 4 patients with biopsy-proven cardiac amyloidosis undergoing treatment | Mayo Clinic AI-ECG amyloid score followed longitudinally during therapy | Serial AI-ECG scores were compared with cardiac biomarkers and echocardiographic findings during treatment | No external validation; case series only | Cardiac amyloidosis, including AL-CA treatment monitoring context | Proof-of-concept case series suggesting that an AI-ECG score originally developed for detection may behave as a longitudinal disease-activity marker; not sufficient to define response thresholds or clinical decision rules |
| Spielvogel et al. (2025) | Serial 99mTc-DPD SPECT/CT | 45 ATTR-CM patients receiving disease- | AI-driven segmentation and quantification of 26 | 17 of 26 AI-extracted SPECT/CT markers decreased after treatment; selected | Single-center longitudinal cohort; no val- | ATTR-CM treatment-associated change | Most direct AI nuclear-imaging evidence that serial quantitative DPD SPECT/CT markers can |
| | | modifying therapy, including 37 ATTRwt-CM and 8 ATTRv-CM | SPECT/CT markers, including SUV metrics, retention index, amyloid-affected volume, and amyloid activity | changes were associated with NT-proBNP, NYHA class, and death or heart-failure hospitalization | idated response threshold | | capture treatment-associated change in ATTR-CM; supports feasibility but does not yet establish standardized response criteria |
| Miller et al. (2026) | Serial 99mTc-PYP SPECT/CT | Patients with cardiac amyloidosis who underwent at least two 99mTc-PYP studies | AI-driven longitudinal PYP quantification using deep learning-derived segmentation and biomarker extraction | Serial cardiac pyrophosphate activity and volume of involvement were assessed against multimodality imaging parameters and outcomes; treated patients showed reductions in VOI and CPA, and increasing CPA was associated with worse outcomes | Retrospective longitudinal validation | ATTR-CM / cardiac amyloidosis with serial PYP imaging | Extends AI SPECT/CT quantification from single-time-point disease-burden measurement to longitudinal disease tracking and outcome-linked change |
| Venneri et al. (2025) | Serial AI-derived echocardiography | 752 ATTR-CM patients with baseline and 12-month echocardiography | Fully automated AI echocardiographic measurement of LVOT-VTI as a surrogate of stroke volume | A ≥5% decrease in AI-derived LVOT-VTI over 12 months was independently associated with all-cause mortality; adjusted HR 1.41, 95% CI 1.13–1.76 | Retrospective longitudinal cohort; not a treatment-response validation study | ATTR-CM disease progression | Supports AI-derived echocardiographic functional markers for disease-progression monitoring, but should not be described as direct treatment-response validation |
| Oikonomou et al. (2025) | Longitudinal ECG and echocardiography | 7,352 TTEs and 32,205 ECGs across Yale New Haven Health System and Houston Methodist | Deep learning on ECG images and TTE videos with longitudinal AI probability trajectories | AI probabilities diverged up to 3 years before nuclear ATTR-CM testing; double-negative sensitivity about 91% and 86%; double-positive specificity about 86% and 89% across cohorts | Two-health-system retrospective longitudinal validation | ATTR-CM preclinical trajectory / risk tracking | Related longitudinal AI evidence, but not a treatment-response study; useful conceptually because it treats AI outputs as trajectories rather than one-time diagnostic labels |

## 2. Feature representations used in cardiac amyloidosis AI studies and their clinical interpretation

| Feature group | Specific features used | Modality / task | Clinical interpretation | Main impact |
|---|---|---|---|---|
| Latent deep learning features | Learned ECG, echo-video, CMR, and scintigraphy representations | ECG, echo, CMR, cardiac scintigraphy with bone-avid tracers | Captures hidden waveform, motion, texture, or uptake patterns not manually defined | Strong discrimination, but limited direct interpretability [14,15,18,19,23,24,27,28,41,44] |
| Cine-CMR radiomics | GLevNonU, GLCM contrast, difference variance, sum average, sum variance, angular second moment, inverse difference moment | CA vs HCM detection | Measures myocardial texture heterogeneity on cine images | Added value beyond wall thickness, LV mass, asymmetric septal hypertrophy, and LGE presence [45] |
| LGE-CMR radiomics | Global, basal, mid-ventricular, and apical myocardial radiomic scores | Diagnosis and prognosis | Reflects LGE texture, regional disease pattern, and myocardial involvement | Added diagnostic/prognostic information beyond conventional staging and imaging assessment [66,77] |
| CT radiomics | Shape/size, first-order intensity, and texture features from LV wall | CA vs aortic stenosis or diffuse myocardial thickening | Captures myocardial attenuation heterogeneity and tissue-pattern differences | May help distinguish amyloid-related myocardial changes from pressure-overload hypertrophy or diffuse myocardial thickening [26,68] |
| SPECT/CT radiomics | First-order, shape, texture, and wavelet-transformed features from SPECT/CT | CA detection | Captures uptake heterogeneity and | May improve classification beyond visual or simple quantitative metrics, but feature-level interpretation remains preliminary [64]. |
| | | | myocardial/CT texture beyond visual grading | |
| Conventional clinical/imaging variables | LV wall thickness, LV mass, asymmetric septal hypertrophy, LGE presence, NT-proBNP, NYHA class, LVEF, LVDD | CMR, CT, echo, clinical models | Familiar clinical markers of hypertrophy, function, fibrosis/infiltration, and disease severity | Anchors AI/radiomics models to clinically meaningful disease features [45,66,68] |
| Quantitative nuclear biomarkers | SUVmax, SUVpeak, TBR, VOI, CPV, CPA | DPD/PYP SPECT/CT quantification | Measures uptake intensity, spatial extent, and composite tracer burden | Most directly interpretable feature group; useful for diagnosis, prognosis, and longitudinal monitoring [21,30,46,48–50] |